\documentclass[aps,prd,onecolumn,groupedaddress,showpacs,nofootinbib,amssymb
]{revtex4}
\usepackage[dvips]{graphicx}
\usepackage{amssymb}
\usepackage{amsmath}
\usepackage{graphicx}
\usepackage{amsfonts}
\usepackage{bm}

\begin{document}

\title{A Study of Finite-time Singularities of Loop Quantum Cosmology Interacting Multifluids}
\author{S.D. Odintsov,$^{1,2,3}$\,\thanks{odintsov@ieec.uab.es}
V.K. Oikonomou,$^{4,5,6}$\,\thanks{v.k.oikonomou1979@gmail.com}}
\affiliation{$^{1)}$ ICREA, Passeig Luis Companys, 23, 08010 Barcelona, Spain\\
$^{2)}$ Institute of Space Sciences (ICE,CSIC) C. Can Magrans s/n,
08193 Barcelona, Spain\\
$^{3)}$ Institute of Space Sciences of Catalonia (IEEC),
Barcelona, Spain\\
$^{4)}$ Department of Physics, Aristotle University of Thessaloniki, Thessaloniki 54124, Greece\\
$^{5)}$ Laboratory for Theoretical Cosmology, Tomsk State University
of Control Systems
and Radioelectronics (TUSUR), 634050 Tomsk, Russia\\
$^{6)}$ Tomsk State Pedagogical University, 634061 Tomsk, Russia
}
\tolerance=5000

\begin{abstract}
In this work we study the occurrence of finite-time cosmological singularities in a cosmological system comprising from three fluids. Particularly, the system contains two dark fluids, namely that of dark energy and dark matter, which are interacting, and of a non-interacting baryonic fluid. For the study we adopt the phase space approach by constructing the cosmological dynamical system in such a way so that it rendered to be an autonomous polynomial dynamical system, and in order to achieve this, we appropriately choose the variables of the dynamical system. By employing a rigid mathematical framework, that of dominant balances analysis, we demonstrate that there exist non-singular solutions of the dynamical system, which correspond to a general set of initial conditions, which proves that no Big Rip or Type III finite-time singularities occur in this LQC multifluid dynamical system. Thus the new feature of this work is that we are able to do this using an analytic technique instead of adopting a numerical approach. In addition, we perform a fixed point analysis of the cosmological dynamical system, and we examine the behavior of the total effective equation of state parameter, at the fixed points, as a function of the free parameters of the system. Finally, we investigate the phenomenological implications of the dark energy equation of state which we assumed that it governs the dark energy fluid.
\end{abstract}


\maketitle

\section{Introduction}

Singularities in cosmology occur in various theoretical contexts, like for example modified gravity \cite{reviews1,reviews2,reviews3,reviews4,reviews5,reviews6} or even multifluid cosmology \cite{LQC1,LQC3,LQC4,LQC5,Salo:2016dsr,Xiong:2007cn,Amoros:2014tha,Cai:2014zga,deHaro:2014kxa}. The definition of a singular point in cosmology was given some time ago from Hawking and Penrose \cite{hawkingpenrose}, and most of the theorems proven by them make use of the null energy condition and also of the facts that at a singular point of the spacetime, geodesics incompleteness occurs and also the curvature scalars diverge. Although in modified gravity the null energy condition may be different in general in comparison to the Einstein-Hilbert case (see for example \cite{Santos:2007bs}), it is generally accepted that the geodesic incompleteness and also that the divergence of the curvature invariants, strongly indicate the presence of a crushing singularity. The singularities in cosmology vary in their effects, and a complete classification of these was performed in \cite{Nojiri:2005sx}. In any case, the singularities to our opinion could be viewed as alternative physics windows in our classical world and have an intrinsic and appealing interest. In particular, finite-time singularities and especially the crushing types, could be viewed as either flaws or shortcomings of the classical theory, or even a doorway to the quantum description of general relativity (especially the initial singularity). This is due to the fact that these cannot be dressed in a similar way to the spacelike singularities of black holes for instance, so their presence casts uncertainty to the predictions of a classical gravitational theory. Thus in general, although singular spacetimes have some interest, due to the fact that closed timelike curves can be ``absorbed'', or more formally be deformed on these, the presence of singularities indicates the inability of the theoretical framework to capture the complete physical description of spacetime.

Quantum theories of gravity though offer a remedy to the singularity problems of classical cosmology. One particularly interesting framework is that of Loop Quantum Cosmology (LQC) \cite{LQC1,LQC3,LQC4,LQC5,Salo:2016dsr,Xiong:2007cn,Amoros:2014tha,Cai:2014zga,deHaro:2014kxa}, in which singularities are removed to a great extent, see for example \cite{Sami:2006wj}. In this work we shall investigate a multifluid cosmological system in the context of LQC focusing on the question whether the quantum framework actually removes the finite time singularities. This problem is addressed for the single fluid case in Ref. \cite{Sami:2006wj}, however, the presence of multiple fluids perplexes the mathematical problem to a great extent. To this end, we shall appropriately form the cosmological equations in such a way, so that these form an autonomous dynamical system of polynomial form, and we shall use a powerful theorem from dynamical systems \cite{goriely}. In this way we shall be able to answer in a firm way whether LQC actually makes the resulting theory singularity-free.

The study of multifluids is in general a quite fertile ground for theoretical modelling of our Universe's evolution, since at present time, the driving force of the accelerated expansion of our Universe, and also of the nature of dark matter, is still unknown. With regard to dark matter, speculations of it's particle nature exist for many years \cite{Oikonomou:2006mh}, however, no direct evidence is given for the moment favoring the existence of a Weakly Interacting Massive Particle, although future experiments might reveal such possibility, see \cite{Oikonomou:2006mh} for an extensive account of this topic. The dark energy problem though is quite more evolved, and various proposals exist in the literature, among which modified gravity is quite promising \cite{reviews1,reviews2,reviews3,reviews4,reviews5,reviews6}. Another quite promising possibility, is that both dark energy and dark matter are described by interacting fluids. In the literature, the fluid cosmology possibility has been extensively studied, see for example \cite{Barrow:1994nx,Tsagas:1998jm,HipolitoRicaldi:2009je,Gorini:2005nw,Kremer:2003vs,Carturan:2002si,Buchert:2001sa,Hwang:2001fb,Cruz:2011zza,Oikonomou:2017mlk,Brevik:2017msy,Nojiri:2005sr,Capozziello:2006dj,Nojiri:2006zh,Elizalde:2009gx,Elizalde:2017dmu,Brevik:2016kuy,Balakin:2012ee,Zimdahl:1998rx}. This possibility is supported by the fact that the dark energy sector dominates overwhelmingly over the dark matter sector after galaxy formation. Moreover, there are hints that dark matter and dark energy depend on each other, and specifically due to the degeneracy of the dark energy models, it is impossible to measure $\Omega_m$, as was shown in \cite{Kunz:2007rk}. Interacting dark energy-dark matter systems are studied in the literature, see for example Refs. \cite{Gondolo:2002fh,Farrar:2003uw,Cai:2004dk,Bamba:2012cp,Guo:2004xx,Wang:2006qw,Bertolami:2007zm,He:2008tn,Valiviita:2008iv,Jackson:2009mz,Jamil:2009eb,He:2010im,Bolotin:2013jpa,Costa:2013sva,Boehmer:2008av,Li:2010ju,Yang:2017zjs} for an  important stream of works. In this paper we shall work with a mixture of three fluids, two modelling dark energy and dark matter, which shall be assumed to have an interaction, and also a baryonic fluid, which will not interact with the other two fluids. These three fluids shall be required to obey the LQC equations of motion, and in order to study the occurrence of finite-time singularities, we shall introduce some appropriately chosen dimensionless variables, and we shall form the cosmological equations in such a way so that the resulting dynamical system is an autonomous polynomial dynamical system (for works invoking cosmological dynamical systems, see for example \cite{Boehmer:2014vea,Bohmer:2010re,Goheer:2007wu,Leon:2014yua,Leon:2010pu,deSouza:2007zpn,Giacomini:2017yuk,Kofinas:2014aka,Leon:2012mt,Gonzalez:2006cj,Alho:2016gzi,Biswas:2015cva,Muller:2014qja,Mirza:2014nfa,Rippl:1995bg,Ivanov:2011vy,Khurshudyan:2016qox,Boko:2016mwr,Odintsov:2017icc,Odintsov:2017tbc,Oikonomou:2017ppp,Odintsov:2015wwp,Bahamonde:2017ize,Ganiou:2018dta}. The reason for this is the fact that the general analytic study of the three fluid system is quite difficult in general. Three fluid and two fluid models of the Universe in the context of LQC are studied in \cite{Li:2010ju,Xiao:2010cy,Zonunmawia:2017ofc,Chen:2008ca}. Specifically, the framework of the three aforementioned fluids was used in \cite{Li:2010ju}, in the context of a dynamical system analysis, but in our case the dynamical system is completely different, for the purposes of finite-time singularities analysis that will follow. The major outcome of this work is that we shall demonstrate in a formal way that finite-time singularities are absent in LQC multifluid cosmology, at least crushing type singularities. The mathematical framework we shall use was proposed sometime ago in Ref. \cite{goriely}, and we shall refer to it as dominant balance analysis. This framework was used in cosmological context later on in Ref. \cite{barrowcotsakis}. In one of the following sections we shall briefly discuss the main features of this analysis, and we shall use this technique in order to reveal the finite-time singularity structure of the three fluid LQC system. Moreover, we shall study in some detail the phase space of the dynamical system of the three and two fluids case, in order to see the attractors of it, and also their stability. Finally, we investigate the phenomenological viability of the dark energy equation of state by confronting the theory with the observational data.

This paper is organized as follows: In section II we provide a brief introduction to the fundamental features of LQC, by focusing on how the modified Friedmann equation of LQC is obtained. In section III we describe in detail the three fluid cosmological system, and we introduce appropriate variables which will be used to construct an autonomous polynomial dynamical system. After obtaining the cosmological dynamical system, we provide a brief overview of the mathematical framework we shall use in order to study the occurrence of finite-time singularities, and we directly apply the method for the cosmological system at hand. We also discuss the difference between a cosmological singularity and of a dynamical system finite-time singularity. In section IV we investigate which are the fixed points of the cosmological dynamical system, and we study the behavior of the total effective equation of state parameter, as a function of the free variables of the cosmological dynamical system. In section V we discuss in some detail the phenomenological implications of the dark energy equation of state which we assumed that controls this sector. We also compare the singularity behavior of the single fluid Einstein-Hilbert theory with the LQC single fluid theory and with the LQC interacting three fluids theory. Finally, the conclusions follow in the end of the paper.

Before getting into the details of this work, it is worth describing in brief the geometric background which we shall assume to be a flat Friedmann-Robertson-Walker (FRW) metric,
\begin{equation}\label{frw}
ds^2 = - dt^2 + a(t)^2 \sum_{i=1,2,3} \left(dx^i\right)^2\, ,
\end{equation}
where $a(t)$ is the scale factor. Moreover, the Ricci scalar for the above metric is,
\begin{equation}\label{ricciscalaranalytic}
R=6\left (\dot{H}+2H^2 \right )\, ,
\end{equation}
where $H=\frac{\dot{a}}{a}$ is the Hubble rate.


\section{A Brief Review of Loop Quantum Cosmology}

In this section we present some fundamental features of holonomy corrected LQC, in order to have a brief and compact idea of what LQC brings along in a cosmological setting. The main feature of spacetime in the context of LQC is that it has a discrete nature, quantified in the Hamiltonian of the quantum theory in terms of the holonomies $h_j=e^{-\frac{i\lambda\sigma_j}{2}}$, with $\sigma_j$ being the Pauli matrices. By using the holonomies, the LQC Hamiltonian is equal to \cite{abl03,bojowald05},
\begin{equation}\label{hamiltonianlqc}
\mathrm{H}_{LQC}=-\frac{2V}{\gamma^3\lambda^3}\Sigma_{i,j,k}\epsilon^{ijk}\mathrm{Tr}[h_i(\lambda)h_j(\lambda)h_i^{-1}(\lambda)\{h_k^{-1},V\}]+\rho V\, ,
\end{equation}
with $\gamma=0.2375$
being the Barbero-Immirzi parameter. In addition, the parameter $\lambda$, which has dimensions of length, is equal to, $\lambda=\sqrt{\frac{\sqrt{3}}{2}\gamma}=0.3203$ and this value corresponds to the square root of the smallest eigenvalue of the  Loop Quantum Gravity area operator \cite{Singh07}. Moreover, $V$ is the
volume of the spacetime, which for the FRW metric (\ref{frw}) is equal to $V=a^3$ and in addition, $\rho$ stands for the total effective energy density of the Universe. The parameter $\beta$ appearing in the holonomies, is the canonically conjugate variable of $V$, and the Poisson bracket of these two is $\{\beta,V\}=\frac{\gamma}{2}$. By calculating the trace of the Hamiltonian, we obtain \cite{he,dmp},
\begin{equation}\label{hamiltonianlqc2}
\mathrm{H}_{LQC}=-3V\frac{\sin^2(\lambda \beta)}{\gamma^2\lambda^2}+\rho V\, ,
\end{equation}
and in conjunction with the Hamiltonian constraint $\mathrm{H}_{LQC}=0$ we obtain the LQC version of the FRW equation, which is,
\begin{equation}\label{hcqrefre1}
\frac{\sin^2(\lambda \beta)}{\gamma^2\lambda^2}=\frac{\rho}{3}\, .
\end{equation}
In view of the Hamiltonian equation $\dot{V}=\{V,\mathrm{H}_{LQC}\}=-\frac{\gamma}{2}\frac{\partial \mathrm{H}_{LQC}}{\partial \beta}$, then we obtain,
\begin{equation}\label{frwhceqn1}
H=\frac{\sin(\lambda \beta)}{\gamma \lambda}\, ,
\end{equation}
or written differently,
\begin{equation}\label{frwhceqn2}
\beta=\frac{\arcsin (2\lambda \gamma H)}{2\lambda}\, .
\end{equation}
Combining Eqs. (\ref{frwhceqn2}) and (\ref{hcqrefre1}), we get,
\begin{equation}\label{frwhceqn3}
\frac{\sin^2(\lambda \frac{\arcsin (2\lambda \gamma H)}{2\lambda})}{\gamma^2\lambda^2}=\frac{\rho}{3}\, .
\end{equation}
It is then a matter of some algebra to get the LQC version of the Friedmann equation, which is,
\begin{equation}\label{lqcfriedmannequation}
H^2=\frac{\rho}{3}\left( 1-\frac{\rho}{\rho_c}\right)\, .
\end{equation}
The parameter $\rho_c$ is of crucial importance in LQC, since it is the maximum allowed value of the energy density of the Universe, and it is equal to $\rho_c=\frac{3}{\gamma^2\lambda^2}\cong 258$. Practically, this parameter quantifies the quantum effects in the LQC Friedmann equation, so when the limit $\rho_c\to \infty$ is taken, the classical  Friedmann equation is obtained, namely $H^2=\frac{\rho}{3}$. In the following we shall make extensive use of the LQC Friedmann equation, by assuming that the total energy density comprises of three fluids, two of which are interacting. Up to date, the LQC background dynamics is well understood \cite{Martineau:2017sti}. With regard to the perturbations issue, there exist various approaches in LQC, for example the deformed algebra method \cite{Barrau:2014maa} and the dressed metric scenario \cite{Agullo:2012sh}. The main open issues are firstly to reduce the gap with the mother theory \cite{Alesci:2016gub}, secondly to take into account trans-Planckian effects \cite{Martineau:2017tdx} and thirdly to develop a concrete numerical LQC \cite{Diener:2017lde}. These are challenges for LQC, but it a promising theoretical framework so we believe that these issues will be firmly addressed in the near future.

\section{LQC Interacting Multifluids and their Dynamical System}

In this section we shall present the theoretical framework of the gravitational theory we shall assume to control the Universe's evolution. As we already mentioned in the introduction, three fluids are considered present in the Universe, two of which interact and correspond to dark energy and dark matter, and a non-interacting baryonic fluid. Also in order to capture the most general effects on the dark energy sector, we shall add a bulk viscosity term in the dark energy fluid. This will allow us to examine a quite general situation. In the context of LQC, the cosmological equation corresponding to the flat FRW metric of Eq. (\ref{frw}) becomes,
\begin{equation}\label{flateinstein}
H^2=\frac{\kappa^2\rho_{tot}}{3}\left( 1-\frac{\rho_{tot}}{\rho_c}\right)\, ,
\end{equation}
where with $\rho_{tot}$ we denote the total energy density of all the matter fluids present in the Universe, hence it is equal to,
\begin{equation}\label{akyrieksisosi}
\rho_{tot}=\rho_m+\rho_d+\rho_b\, ,
\end{equation}
and $\rho_c$ is the critical density defined below Eq. (\ref{lqcfriedmannequation}). In the above equation $\rho_m$, $\rho_d$ and $\rho_b$ stand for the energy density of dark matter, the energy density of dark energy and the energy density of the baryons respectively. By differentiating Eq. (\ref{flateinstein}) with respect to the cosmic time we get,
\begin{equation}\label{derivativeofh}
\dot{H}=-\frac{\kappa^2}{2}\left(\rho_m+\rho_d+\rho_b+p_{tot}\right)\left( 1-2\frac{\rho_m+\rho_d+\rho_b+p_{tot}}{\rho_c}\right)\, ,
\end{equation}
with $p_{tot}$ denoting the total pressure of the matter fluids and effectively this is equal to the pressure of the dark energy fluid $p_d$, since dark matter and baryonic matter are pressure-less. Also the equation of state (EoS) for the dark energy fluid shall be assumed to be as follows \cite{Nojiri:2005sr},
\begin{equation}\label{darkenergyeos}
p_d=-\rho_d-A\kappa^4\rho_d^2\, ,
\end{equation}
where $A$ is real parameter, which is dimensionless. In a later section we shall examine the phenomenological implications of such a dark energy EoS. From the energy-momentum conservation, we obtain the following equations for the energy densities,
\begin{align}\label{continutiyequations}
& \dot{\rho}_b+3H\rho_b=0\,  \\ \notag &
\dot{\rho}_m+3H\rho_m=Q\, \\ \notag &
\dot{\rho}_d+3H(\rho_d+p_d)=-Q\, ,
\end{align}
with $Q$ denoting the interaction term between the dark sectors of the above fluids. It is conceivable that when $Q>0$, the dark energy sector loses energy, while when $Q<0$, the dark matter fluid loses energy. A phenomenologically interesting form of the interaction term $Q$ is the following \cite{CalderaCabral:2008bx,Pavon:2005yx,Quartin:2008px,Sadjadi:2006qp,Zimdahl:2005bk},
\begin{equation}\label{qtermform}
Q=3H(c_1\rho_m+c_2\rho_d)\, ,
\end{equation}
where $c_1$, $c_2$ must be simultaneously positive or negative, for physical consistency reasons.


Having the gravitational equations of the LQC three-fluid system, namely Eqs. (\ref{flateinstein}) and (\ref{derivativeofh}), we shall now form a dynamical system by appropriately choosing the dimensionless variables in order to obtain an autonomous dynamical system of polynomial form. Then, by using a concrete mathematical framework for polynomial autonomous dynamical systems, developed in \cite{goriely}, we shall investigate whether the dynamical system has singular solutions, that is, whether the variables develop a finite-time singularity. Also we shall investigate whether these finite-time singularities of the dynamical system are also cosmological singularities, and this discrimination between cosmological-physical singularities and of dynamical system singularities must always be done. The theorems we shall use for the autonomous polynomial dynamical system were performed firstly in Ref. \cite{goriely}, and were later applied in a cosmological framework in  Ref. \cite{barrowcotsakis}. At a later section we shall briefly describe this method of Ref. \cite{goriely} in order to maintain the article self-contained. Before we proceed, it is vital to make the discrimination between a dynamical system singularity and of a physical cosmological singularity. The classification of physical cosmological finite-time singularities was firstly performed in Ref. \cite{Nojiri:2005sx}, and the singularities are classified as follows,
\begin{itemize}
\item Type I Singularity (``Big Rip Singularity''): Crushing type singularity, it is a metric singularity. It is a future spacelike singularity, and in this case at a time instance $t=t_s$, the scale factor
$a(t)$, the total energy density $\rho_{\mathrm{eff}}$ and the total
pressure $p_\mathrm{eff}$, diverge. Note that the energy density and the pressure are physical quantities defined on a three dimensional spacelike hypersurface, so these diverge for a Big Rip, and also only for the Big Rip singularity the scale factor diverges at $t=t_s$.
\item Type II Singularity (``Sudden Singularity''):
This is also known as pressure singularity
\cite{barrowsudden,barrowsudden1}, and in this case, only the total effective pressure diverges, while the scale factor and the energy density remain finite at $t=t_s$.
\item Type III Singularity: In this case, the total effective energy density and the total effective pressure diverge at $t=t_s$, while the scale factor remains finite.
\item Type IV Singularity: The most soft from a phenomenological point of view, since all the physical quantities defined on the three dimensional spacelike hypersurface $t=t_s$, are finite, and only the higher derivatives of the Hubble rate diverge at $t=t_s$. The phenomenological implications of this singularity type, were studied in Refs. \cite{Odintsov:2015gba,Odintsov:2015jca,Barrow:2015ora,Nojiri:2015fra,Oikonomou:2015qfh}.
\end{itemize}

The major contribution of this work in the study of cosmological finite-time singularities, is that we will prove analytically that no singular solutions exist for the cosmological dynamical system. In the case that only the dark energy fluid was present, it is easy to show that for the EoS appearing in Eq. (\ref{darkenergyeos}) for a LQC framework, a Type III cosmological singularity occurs, see for example \cite{Sami:2006wj,Nojiri:2005sx}. However, for the three-fluids system it is impossible to show this in an analytic way. This gap will be filled by our results, which prove that LQC actually makes all the singularities disappear. Let us now demonstrate how an autonomous dynamical system of polynomial type can be obtained by combining Eqs. (\ref{flateinstein}), (\ref{derivativeofh}) and (\ref{continutiyequations}). We make the following choice of the dynamical variables,
\begin{equation}\label{variablesofdynamicalsystem}
x_1=\frac{\kappa^2\rho_d}{3H^2},\,\,\,x_2=\frac{\kappa^2\rho_m}{3H^2},\,\,\,x_3=\frac{\kappa^2\rho_b}{3H^2},\,\,\,z=\frac{H^2}{\kappa^2\rho_c}\, .
\end{equation}
Also the variables $x_i$, $i=1,2,3$ and $z$ are constrained for all cosmic times, to obey the Friedmann constraint, which in the case of LQC is the following,
\begin{equation}\label{friedmannconstraint}
x_1+x_2+x_3-z\left(x_1+x_2+x_3\right)^2=1\, .
\end{equation}
The Friedmann constraint (\ref{friedmannconstraint}) will prove to be very crucial for the determination of solutions of the dynamical system. In addition, the total EoS parameter $w_{eff}$ of the three fluids system, is written in terms of the dynamical variables  (\ref{variablesofdynamicalsystem}), in the following way,
\begin{equation}\label{equationofstatetotal}
w_{eff}=-x_1-3Ax_1^2z\, .
\end{equation}
By combining Eqs. (\ref{flateinstein}), (\ref{derivativeofh}) and (\ref{continutiyequations}), for the choice of the dynamical variables (\ref{variablesofdynamicalsystem}), after quite some algebra, we obtain the following dynamical system,
\begin{align}\label{dynamicalsystemmultifluid}
& \frac{\mathrm{d}x_1}{\mathrm{d}N}=-\frac{\kappa^2Q}{3H^3}+9 A x_1^3 z-27 A x_1^2 z+3 w_d x_1^2-3 w_d x_1-18 x_1^3 z+3 x_1^2+3 x_1 x_2+3 x_1 x_3-3 x_1\\ \notag &
-18 w_d x_1^3 z-18 w_d x_1^2 x_2 z-36 x_1^2 x_2 z-36 x_1^2 x_3 z-18 x_1 x_2^2 z\\ \notag &
-54 A x_1^4 z^2-54 A x_1^3 x_2 z^2-54 A x_1^3 x_3 z^2-18 w_d x_1^2 x_3 z-36 x_1 x_2 x_3 z-18 x_1 x_3^2 z\, ,
\\ \notag &
\frac{\mathrm{d}x_2}{\mathrm{d}N}=\frac{\kappa^2Q}{3H^3}+9 A x_1^2 x_2 z+3 w_d x_1 x_2-18 x_1^2 x_2 z+3 x_1 x_2+3 x_2^2+3 x_2 x_3-3 x_2 \\ \notag &
-18 w_d x_1^2 x_2 z-18 w_d x_1 x_2^2 z-36 x_1 x_2^2 z-36 x_1 x_2 x_3 z-18 x_2^3 z\\ \notag & -54 A x_1^3 x_2 z^2-54 A x_1^2 x_2^2 z^2-54 A x_1^2 x_2 x_3 z^2-18 w_d x_1 x_2 x_3 z-36 x_2^2 x_3 z-18 x_2 x_3^2 z\, , \\ \notag &
\frac{\mathrm{d}x_3}{\mathrm{d}N}=9 A x_1^2 x_3 z-18 w_d x_1^2 x_3 z+3 w_d x_1 x_3-18 x_1^2 x_3 z+3 x_1 x_3+3 x_2 x_3+3 x_3^2-3 x_3\\ \notag &
-18 w_d x_1 x_2 x_3 z-18 w_d x_1 x_3^2 z-36 x_1 x_2 x_3 z-36 x_1 x_3^2 z-18 x_2^2 x_3 z-36 x_2 x_3^2 z\\ \notag & -54 A x_1^3 x_3 z^2-54 A x_1^2 x_2 x_3 z^2-54 A x_1^2 x_3^2 z^2-18 x_3^3 z
\,  \\ \notag &
\frac{\mathrm{d}z}{\mathrm{d}N}=-9 A x_1^2 z^2+18 w_d x_1^2 z^2-3 w_d x_1 z+18 x_1^2 z^2-3 x_1 z-3 x_2 z-3 x_3 z\\ \notag &
18 w_d x_1 x_2 z^2+18 w_d x_1 x_3 z^2++36 x_1 x_2 z^2+36 x_1 x_3 z^2+18 x_2^2 z^2\\ \notag & 54 A x_1^3 z^3+54 A x_1^2 x_2 z^3+54 A x_1^2 x_3 z^3++36 x_2 x_3 z^2+18 x_3^2 z^2\, ,
\end{align}
where instead of the cosmic time, we used the $e$-foldings number $N$. Also, for $Q$ chosen as in Eq. (\ref{qtermform}), the terms containing $Q$ in the dynamical system (\ref{dynamicalsystemmultifluid}), become in terms of the variables $x_1$ and $x_2$ as follows,
\begin{equation}\label{additionalterms}
\frac{\kappa^2Q}{3H^3}=3c_1x_2+3c_2x_1\, .
\end{equation}
In the following we shall investigate the finite-time singularity structure of the above dynamical system, by also discriminating physical from dynamical system singularities. This analysis shall be performed in the next section.

\subsection{Dominant-Balance Singularity Analysis of Autonomous Dynamical Systems}

Having obtained a polynomial type autonomous dynamical system corresponding to the three fluid dynamical system of the previous section, we now proceed to the concrete singularity structure analysis of the solutions. Our aim is to investigate whether general singular solutions exist, that is, if some of the solutions $x_1(N)$, $x_2(N)$, $x_3(N)$ and $z(N)$ become singular at some finite-time instance, for general initial conditions. It is important to note that the condition for having solutions for general initial conditions is compelling. Before we discuss the physical significance of the singular solutions, if these exist, we shall briefly review the dominant balance analysis of Ref. \cite{goriely}, in order  to render the article self-contained.

Particularly, in Ref. \cite{goriely}, the authors provided sufficient conditions for the finite-time singularity occurrence of a polynomial autonomous dynamical system, of any dimension. Hereafter we shall name this theoretical framework ``dominant balance analysis''. The method is particularly simple and rigid, consider for example an autonomous $n-$dimensional dynamical system of the form,
\begin{equation}\label{dynamicalsystemdombalanceintro}
\dot{x}=f(x)\, ,
\end{equation}
where $x$ is a real vector of $R^n$, and $f(x)=\left(f_1(x),f_2(x),...,f_n(x)\right )$ is some real polynomial vector. A finite-time singularity of the dynamical system above, is formally a singularity that depends on the initial conditions chosen for the dynamical system. To be specific, the singular solution will in general have the form $(t-t_c)^{-p}$, where $t_c$ is some arbitrary constant determined by the initial conditions. Consider for example the one dimensional dynamical system $\frac{\mathrm{d}y}{\mathrm{d}x}=\frac{1}{x^2y^2}$, which has the solution $y=(\frac{1}{x}-c)^{-1}$, with $c$ being an integration constant.  Obviously, the solution is singular at  $\frac{1}{x}=c$, so this singularity is a moving singularity, which strongly depends on the initial conditions. The method is simple and consists of the following steps:

\begin{itemize}

\item In order to determine the existence of general solutions of the dynamical system that may become singular at finite-time, it is required to find truncations (decompositions) of the function $f(x)$ appearing in Eq. (\ref{dynamicalsystemdombalanceintro}), in dominant and subdominant parts. Obviously, the behavior near the singularities is controlled by the dominant truncation, denoted as $\hat{f}(x)$, and the dynamical system reduces to,
\begin{equation}\label{dominantdynamicalsystem}
\dot{x}=\hat{f}(x)\, .
\end{equation}
In principle, there are multiple ways to find dominant truncations, but the theorem may work even for one of these, once the conditions that ensure a general solution, are satisfied. Also, the vector $\hat{f}(x)$ is constructed by using a single polynomial term for each of it's $n$-dimensional entries. Now for each $x_i$, $i=1,2,...,n$ of the vector $x$, we write,
\begin{equation}\label{decompositionofxi}
x_1(\tau)=a_1\tau^{p_1},\,\,\,x_2(t)=a_2\tau^{p_2},\,\,\,....,x_n(t)=a_n\tau^{p_n}\, ,
\end{equation}
and it is required that the  solution $x$ may be written in $\psi$-series in terms of the parameter $\tau=t-t_c$, where $t_c$ stands for the singularity time instance. The rest of the method is very simple conceptually, since by substituting the $x_i$'s appearing in Eq. (\ref{decompositionofxi}) in Eq. (\ref{dominantdynamicalsystem}), for each different entry of $\hat{f}$ corresponding to each $x_i$, one should simply equate the powers of the resulting polynomials. In effect, this determines the parameters $p_i$, $i=1,2,...,n$, with the first constraint in the solutions being that only fractional numbers or integers numbers are allowed for the $p_i$'s. By using the found $p_i$'s, the vector $\vec{p}=(p_1,p_2,...,p_n)$ is formed, which has a significance in the rest of the method. Having found the $p_i$'s, by equating the coefficients of the polynomials in Eqs. (\ref{dominantdynamicalsystem}), (\ref{decompositionofxi}), the coefficients $a_i$ are determined, and then the vector $\vec{a}=(a_1,a_2,a_3,....,a_n)$ can be constructed, which is called dominant balance. The constraint here is that only non-zero dominant balances are allowed, so we form the dominant balance $(\vec{a},\vec{p})$.

\item The rest of the method-theorem of Ref. \cite{goriely} is simple matrix algebra. If the dominant balance $\vec{a}=(a_1,a_2,a_3,....,a_n)$ is complex for some $a_i$'s, or even one of these, then the dynamical system has no finite-time singularities, and it has if $\vec{a}=(a_1,a_2,a_3,....,a_n)$ is real, see \cite{goriely} for the proof and details on this.

\item What no remains is to ensure the existence of general initial solutions that may or may not be singular. There is a rigid way to determine the existence of general solutions, by calculating the Kovalevskaya matrix $R$,
\begin{equation}\label{kovaleskaya}
R=\left(%
\begin{array}{ccccc}
  \frac{\partial \hat{f}_1}{\partial x_1} & \frac{\partial \hat{f}_1}{\partial x_2} & \frac{\partial \hat{f}_1}{\partial x_3} & ... & \frac{\partial \hat{f}_1}{\partial x_n} \\
  \frac{\partial \hat{f}_2}{\partial x_1} & \frac{\partial \hat{f}_2}{\partial x_2} & \frac{\partial \hat{f}_2}{\partial x_3} & ... & \frac{\partial \hat{f}_2}{\partial x_n} \\
  \frac{\partial \hat{f}_3}{\partial x_1} & \frac{\partial \hat{f}_3}{\partial x_2} & \frac{\partial \hat{f}_3}{\partial x_3} & ... & \frac{\partial \hat{f}_3}{\partial x_n} \\
  \vdots & \vdots & \vdots & \ddots & \vdots \\
  \frac{\partial \hat{f}_n}{\partial x_1} & \frac{\partial \hat{f}_n}{\partial x_2} & \frac{\partial \hat{f}_n}{\partial x_3} & ... & \frac{\partial \hat{f}_n}{\partial x_n} \\
\end{array}%
\right)-\left(%
\begin{array}{ccccc}
  p_1 & 0 & 0 & \cdots & 0 \\
  0 & p_2 & 0 & \cdots & 0 \\
  0 & 0 & p_3 & \cdots & 0 \\
  \vdots & \vdots & \vdots & \ddots & 0 \\
  0 & 0 & 0 & \cdots & p_n \\
\end{array}%
\right)\, ,
\end{equation}
at the non-zero balance $\vec{a}$ found in the previous step. If the method is applied in a correct way, the eigenvalues of the Kovalevskaya matrix $R(\vec{a})$ are required to have the form $(-1,r_2,r_3,...,r_{n})$.

\item Now the generality of the solutions is ensured if all the eigenvalues $r_2,r_3,...,r_{n}$ are positive. Hence if we found previously that $\vec{a}$ is real for all it's entries, and also $r_i>0$, $i=2,3,...,n$, then general singular solutions exist. If the eigenvalues are positive, and also $\vec{a}$ is complex for some entries, then no singular solutions exist, and hence no general initial conditions may lead to singular solutions for the $x_i$'s. In all other cases, the solutions are degenerate or a small set of initial conditions leads to these.

\item Finally, if a general singular is found, then the singularity occurs in the orthant of the  $x_i$ configuration space corresponding to $a_i$. For example if a singular behavior is found,  and $a_2<0$, then the singularity occurs for negative values of $x_2$.

\end{itemize}

In the next subsection we shall apply this method for the dynamical system (\ref{dynamicalsystemmultifluid}). Before getting into the details of this analysis, it is vital at this point to discuss an important issue, namely the difference between a physical cosmological finite-time and a dynamical system finite-time singularity. Basically speaking, this crucially depends on the choice of the variables of the dynamical system. In our case, for the variables chosen as in Eq. (\ref{variablesofdynamicalsystem}), a dynamical system singularity in the parameter $z$ clearly indicates either a Big Rip singularity or a Type III physical singularity, due to the fact that the total energy density would diverge. Hence in this case, a crushing type singularity is possibly underlying the dynamical system. Now a singularity in the variables $x_i$ crucially depends on the fact if the parameter $z\sim H^2$ is singular, in which case the energy densities would possibly diverge. The general case can be quite complicated, and a full account on this delicate issue will be given elsewhere, for the Einstein-Hilbert case of the three-fluid system. However, it will prove that in our case such an analysis is not required, since the dynamical system certainly does not contain finite-time singularities, as we now demonstrate.

\subsection{Dominant Balance Analysis of the three-fluid Cosmological Dynamical System}

Let us now apply the method of the previous section to the dynamical system (\ref{dynamicalsystemmultifluid}), in order to investigate when the singularities do not occur, or equivalently, when general non-singular solutions occur. As we demonstrate, the general solutions of the dynamical system are non-singular, or to put formally, there exist general initial conditions that lead to non-singular solutions of the dynamical system. Of course there exist limited sets that may lead to singular solutions, but as we stated, these are limited sets, so of limited interest. Our main interest will be on solutions originating from general initial conditions.

Since the dynamical system (\ref{dynamicalsystemmultifluid}) is expressed in terms of the $e$-foldings number $N$, we assume that the variables $x_1(N)$, $x_2(N)$, $x_3(N)$ and $z(N)$ near the singularities take the following form at leading order,
\begin{equation}\label{decompositionofxiactualexample}
x_1(N)=a_1(N-N_c)^{p_1},\,\,\,x_2(N)=a_2(N-N_c)^{p_2},\,\,\, x_3(N)=a_3(N-N_c)^{p_3},\,\,\,z(N)=a_4(N-N_c)^{p_4}\, ,
\end{equation}
and in effect we look for balances of the form $(\vec{a},\vec{p})$, with the vectors $\vec{a}$ and $\vec{p}$ having the following form,
\begin{equation}\label{balancesactualcase}
\vec{a}=(a_1,a_2,a_3,a_4),\,\,\,\vec{p}=(p_1,p_2,p_3,p_4)\, .
\end{equation}
We can write the dynamical system (\ref{dynamicalsystemmultifluid})  as $\frac{\mathrm{d}\vec{x}}{\mathrm{d}N}=f(\vec{x})$, where the vector $\vec{x}$ has the form $\vec{x}=(x_1,x_2,x_3,z)$, and also the vector function $f(x_1,x_2,x_3,z)$ is equal to,
\begin{equation}\label{functionfmultifluid}
f(x_1,x_2,x_3,z)=\left(%
\begin{array}{c}
 f_1(x_1,x_2,x_3,z) \\
  f_2(x_1,x_2,x_3,z) \\
   f_3(x_1,x_2,x_3,z) \\
   f_4(x_1,x_2,x_3,z) \\
\end{array}%
\right)\, ,
\end{equation}
where the functions $f_i(x_1,x_2,x_3)$, $i=1,2,3,4$ are given below,
\begin{align}\label{functionsfi}
& f_1(x_1,x_2,x_3,z)= 3c_1x_2+3c_2x_1+9 A x_1^3 z-27 A x_1^2 z+3 w_d x_1^2-3 w_d x_1-18 x_1^3 z+3 x_1^2+3 x_1 x_2+3 x_1 x_3-3 x_1\\ \notag &
-18 w_d x_1^3 z-18 w_d x_1^2 x_2 z-36 x_1^2 x_2 z-36 x_1^2 x_3 z-18 x_1 x_2^2 z\\ \notag &
-54 A x_1^4 z^2-54 A x_1^3 x_2 z^2-54 A x_1^3 x_3 z^2-18 w_d x_1^2 x_3 z-36 x_1 x_2 x_3 z-18 x_1 x_3^2 z\, , \\ \notag &
f_2(x_1,x_2,x_3,z)=3c_1x_2+3c_2x_1+9 A x_1^2 x_2 z+3 w_d x_1 x_2-18 x_1^2 x_2 z+3 x_1 x_2+3 x_2^2+3 x_2 x_3-3 x_2 \\ \notag &
-18 w_d x_1^2 x_2 z-18 w_d x_1 x_2^2 z-36 x_1 x_2^2 z-36 x_1 x_2 x_3 z-18 x_2^3 z\\ \notag & -54 A x_1^3 x_2 z^2-54 A x_1^2 x_2^2 z^2-54 A x_1^2 x_2 x_3 z^2-18 w_d x_1 x_2 x_3 z-36 x_2^2 x_3 z-18 x_2 x_3^2 z\, , \\ \notag &
f_3(x_1,x_2,x_3,z)=9 A x_1^2 x_3 z-18 w_d x_1^2 x_3 z+3 w_d x_1 x_3-18 x_1^2 x_3 z+3 x_1 x_3+3 x_2 x_3+3 x_3^2-3 x_3\\ \notag &
-18 w_d x_1 x_2 x_3 z-18 w_d x_1 x_3^2 z-36 x_1 x_2 x_3 z-36 x_1 x_3^2 z-18 x_2^2 x_3 z-36 x_2 x_3^2 z\\ \notag & -54 A x_1^3 x_3 z^2-54 A x_1^2 x_2 x_3 z^2-54 A x_1^2 x_3^2 z^2-18 x_3^3 z
\,  \\ \notag &
f_4(x_1,x_2,x_3,z)=-9 A x_1^2 z^2+18 w_d x_1^2 z^2-3 w_d x_1 z+18 x_1^2 z^2-3 x_1 z-3 x_2 z-3 x_3 z\\ \notag &
18 w_d x_1 x_2 z^2+18 w_d x_1 x_3 z^2++36 x_1 x_2 z^2+36 x_1 x_3 z^2+18 x_2^2 z^2\\ \notag & 54 A x_1^3 z^3+54 A x_1^2 x_2 z^3+54 A x_1^2 x_3 z^3++36 x_2 x_3 z^2+18 x_3^2 z^2\, .
\end{align}
What now remains is to find truncations of the vector function $f(x_1,x_2,x_3,z)$ of Eq. (\ref{functionfmultifluid}), which will indicate wether general non-singular solutions exist.

In principle there are many self-consistent possible truncations of the vector function $f(x_1,x_2,x_3,z)$, one of which is,
\begin{equation}\label{truncation1}
\hat{f}(x_1,x_2,x_3,z)=\left(
\begin{array}{c}
 3 x_1(N) x_2(N) \\
 3 w_d x_1(N) x_2(N)+3 x_1(N) x_2(N) \\
 -54 A x_1(N)^2 x_3(N)^2 z(N)^2 \\
 54 A x_1(N)^2 z(N)^3 x_2(N) \\
\end{array}
\right)\, .
\end{equation}
Then by applying the method of the previous section, we easily obtain the $\vec{p}$, which is,
\begin{equation}\label{vecp1}
\vec{p}=( -1, -1, -1, 1 )\, .
\end{equation}
Accordingly, for the obtained solution $\vec{p}$ being as above, the only non-zero vector-solution $\vec{a}$ is equal to,
\begin{align}\label{balancesdetails1}
& \vec{a}_1=\Big{(}-\frac{1}{3 (w_d+1)}, -\frac{1}{3}, -\frac{1}{3}, -\frac{\sqrt{-w_d^2-2 w_d-1}}{\sqrt{2} \sqrt{A}}\Big{)} \, .
\end{align}
At this point let us investigate when the vector $\vec{a}$ is complex, and it is easy to see that when $A<0$, it is always real, and when $A>0$, it is always complex. Then according to the theorems we discussed in the previous section, when $A>0$, there are non-singular solutions, and when $A<0$ there exist singular solutions. Also note that the case $A=0$ leads to inconsistencies, since the $\vec{a}$ blows up, so no concrete conclusion can be obtained.

Now it remains to check if the singular or non-singular solutions are general, that is, if these originate from general initial conditions. The answer to this question can be obtained by calculating the Kovalevskaya matrix on $\vec{a}$. By doing this, we obtain,
\begin{equation}\label{kobvalev1}
R_1=\left(
\begin{array}{cccc}
 0 & -\frac{1}{w_d+1} & 0 & 0 \\
 -w_d-1 & -\frac{w_d}{w_d+1}-\frac{1}{w_d+1}+1 & 0 & 0 \\
 \frac{2 \left(w_d^2+2 w_d+1\right)}{w_d+1} & 0 & \frac{2 \left(w_d^2+2 w_d+1\right)}{(w_d+1)^2}+1 & \frac{2 \sqrt{2} \sqrt{A} \sqrt{w_d^2+2 w_d+1}}{3 (w_d+1)^2} \\
 \frac{2 \left(w_d^2+2 w_d+1\right)}{w_d+1} & 0 & \frac{2 \left(w_d^2+2 w_d+1\right)}{(w_d+1)^2} & \frac{2 \sqrt{2} \sqrt{A} \sqrt{w_d^2+2 w_d+1}}{3 (w_d+1)^2}-1 \\
\end{array}
\right)\, ,
\end{equation}

and the corresponding eigenvalues can be easily calculated, and these are,
\begin{equation}\label{eigenvalues1}
(r_1,r_2,r_3,r_4)=(-1,1,r_,r_+)\, ,
\end{equation}
where $r_{\pm}$ are equal to,
\begin{equation}\label{rpm}
r_{\pm}=\frac{\sqrt{2} \sqrt{A} \sqrt{(w_d+1)^2}\pm \sqrt{2} \sqrt{(w_d+1)^2 \left(A+18 (w_d+1)^2\right)}+3 w_d (w_d+2)+3}{3 (w_d+1)^2}\, .
\end{equation}
It is obvious that a general conclusion can be obtained only for $A>0$, and it can be checked that when $A$ and $w_d$ satisfy the following inequality,
\begin{equation}\label{inequalityAwd}
\sqrt{2} \sqrt{A} \sqrt{(w_d+1)^2}+3 w_d (w_d+2)+3>\sqrt{2} \sqrt{(w_d+1)^2 \left(A+18 (w_d+1)^2\right)}\, ,
\end{equation}
the eigenvalues (\ref{eigenvalues1}) are always positive, for all positive $A$. Hence, in this case, there exist non-singular solutions of the dynamical system, which are general and therefore correspond to general initial conditions. When the inequality does not hold true, then the eigenvalue $r_-$ becomes negative, and hence the solutions of the dynamical are not general and correspond to a limited set of initial conditions. Therefore, we demonstrated that for the complicated system of three fluids, the LQC framework leads to solutions which do not develop a finite-time singularity for a quite general set of initial conditions. Actually, our analysis for the variables chosen as in Eq. (\ref{variablesofdynamicalsystem}), clearly demonstrate the absence of all Types I, Type II and Type III singularities. For the Type III and Type I case, this is clear since the variable $z$ never diverges, it is non-singular. Also for the Type II case, the same argument applies, since $H^2$ is finite, and from Eq. (\ref{darkenergyeos}) it can be seen that the pressure can never diverge. Actually, the pressure of the cosmological system corresponds to the pressure of the dark energy fluid, since the rest of the fluids are pressureless. As for the Type IV case, the fact that the variable $z$ and also $\dot{H}$ are non-singular (see the expression (\ref{derivativeofh})), does not guarantee the finiteness of the higher derivatives of the Hubble rate. Hence it might be possible that Type IV singularities might occur in the system, but with the theoretical framework we used, it is not possible to see this, at least in the dynamical system level.

Thus as in most cases related to LQC, in this case too, the LQC framework completely erases the singularity occurrence in the theory, at least the crushing and pressure singularities. It is worth having a qualitative idea of the phase space of the three fluids, and in the next section we will study the fixed points of the cosmological system under study.

\section{Fixed Point Analysis of the LQC Multifluid Cosmological Model and Behavior of the Equation of State}

In this section we shall find the fixed points of the cosmological system of the three fluids and we shall investigate their stability. At the end of the section we shall investigate the behavior of the total EoS parameter, given in Eq. (\ref{equationofstatetotal}), as a function of the free parameters of the theory, namely the EoS parameter of the dark energy and of $c_1$, $c_2$. The fixed points of the dynamical system can be found using standard techniques of dynamical systems,  and a complete answer on the stability of fixed points can be found by using the linearization technique \cite{jost,wiggins,voyatzis}. If the fixed point is hyperbolic, that is, if the eigenvalues of the linearization matrix have non-zero real parts, the stability is easy to be determined. If all the eigenvalues have negative real part, the fixed point is stable, and if one eigenvalue is positive, then the fixed point is unstable. Let us denote the fixed points of the dynamical system (\ref{dynamicalsystemmultifluid}) with $\phi_*$,  and also the Jacobian matrix of the linearized dynamical system near at each fixed point as $\mathcal{J}(g)$, which is equal to,
\begin{equation}\label{jaconiab}
\mathcal{J}=\sum_i\sum_j\Big{[}\frac{\mathrm{\partial f_i}}{\partial
x_j}\Big{]}\, .
\end{equation}
By solving the equation $f(x_1,x_2,x_3,z)=0$, where the function $f(x_1,x_2,x_3,z)$ appears in Eq. (\ref{functionfmultifluid}),  we obtain the following fixed points for the dynamical system (\ref{dynamicalsystemmultifluid}),
\begin{align}\label{fixedpointsc20}
& \phi_1^*=\{x_1\to 0,x_2\to 0,x_3\to 0\}, \\ \notag &
\phi_2^*=\{x_1\to 0,x_2\to 0,x_3\to 0,z\to 0\},\\ \notag &
\phi_3^*=\{x_1\to 0,x_2\to 0,x_3\to 1,z\to 0\},\\ \notag &
\phi_4^*=\left\{x_1\to \frac{-\sqrt{(c_1-c_2-w_d)^2+4 c_1 w_d}-c_1+c_2+w_d}{2 w_d},x_2\to \frac{\frac{c_1^2}{w_d}+\frac{c_1 \sqrt{(c_1-c_2-w_d)^2+4 c_1 w_d}}{w_d}-\frac{c_1 c_2}{w_d}+c_1}{2 c_1},x_3\to 0,z\to 0\right\}\\ \notag &
\phi_5^*=\left\{x_1\to \frac{\sqrt{(c_1-c_2-w_d)^2+4 c_1 w_d}-c_1+c_2+w_d}{2 w_d},x_2\to \frac{\frac{c_1^2}{w_d}-\frac{c_1 c_2}{w_d}-\frac{c_1 \sqrt{(c_1-c_2-w_d)^2+4 c_1 w_d}}{w_d}+c_1}{2 c_1},x_3\to 0,z\to 0\right\}\, .
\end{align}
The first two fixed points are not hyperbolic, as it can be checked, and only the fixed points $\phi_3^*$, $\phi_4^*$ and $\phi_5^*$ are hyperbolic, that is, the corresponding Jacobian matrix of the linearized system has eigenvalues with non-zero real parts. We omit the explicit form of the corresponding eigenvalues, since these are quite lengthy to be presented here. A thorough analysis of the parameter space, leads to the general conclusion that the fixed points are unstable, regardless if $w_d$ is positive or negative, and for all the values of $c_1$, $c_2$ (recall that the latter two must have the same sign for physical consistency).

Let us now investigate the behavior of the EoS, by remembering that $A$ must be positive in order to have general solutions for the dynamical system, which are non-singular, as we demonstrated in the previous section. Recall that the total EoS parameter is given in Eq. (\ref{equationofstatetotal}), so let us evaluate it at the values of $x_i$, $i=1,2,3$ and $z$, corresponding to the unstable fixed points. For our numerical analysis we shall fix the value of the dark energy EoS parameter $w_d$ to be a quintessential value, for example $w_d=-0.5$ and also for a phantom value $w_d=-1.5$. In Fig. \ref{plot1} we present the contour plot of the values of the total EoS parameter as a function of $c_1$ and $c_2$ for $w_d=-0.5$ (left plot) and for $w_d=-1.5$ in the ranges $c_1=[-0.5,0.5]$ and $c_2=[-0.5,0.5]$. The values of $w_{eff}$ from left to right in each plot correspond to the range $w_{eff}=[-2,1]$. As it can be seen, various cosmological evolutions may be realized with the three fluid cosmological system, varying from a phantom era, to a quintessential and also to a mater dominated era as a limiting case (the right region in the contour plots).
\begin{figure}[h]
\centering
\includegraphics[width=20pc]{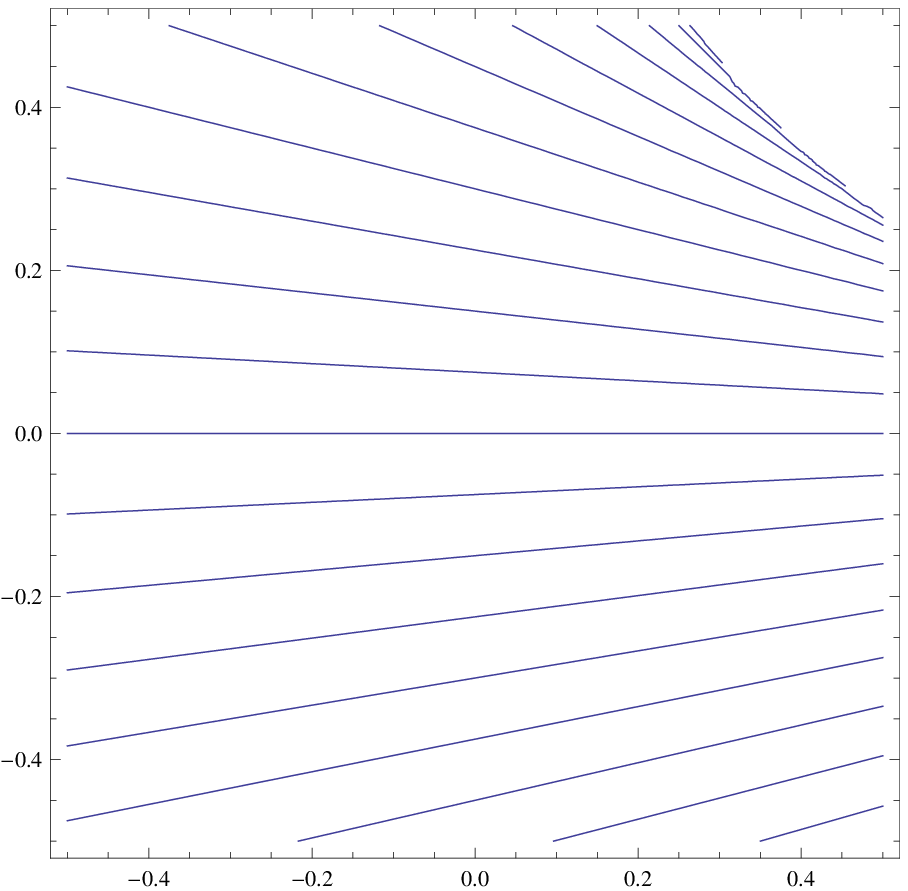}
\includegraphics[width=20pc]{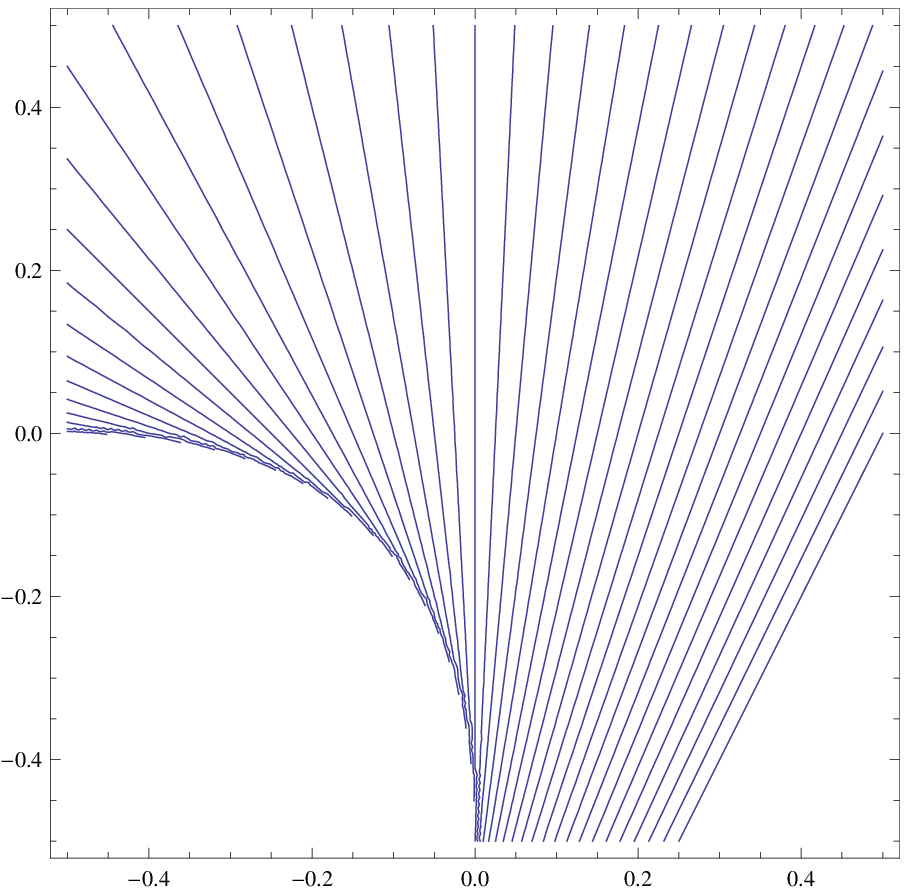}
\caption{{\it{Contour plot of the values of the total EoS parameter $w_{eff}$ as a function of $c_1$ and $c_2$ for $w_d=-0.5$ (left plot) and for $w_d=-1.5$ (right plot) in the ranges $c_1=[-0.5,0.5]$ and $c_2=[-0.5,0.5]$. The values of $w_{eff}$ from left to right in each plot correspond to the range $w_{eff}=[-2,1]$.}}} \label{plot1}
\end{figure}

\section{Physical Implications of the Dark Energy Equation of State}

For the considerations made in the previous sections, we used an EoS for the dark matter fluid which contains quadratic powers of the energy density, namely the one appearing in Eq. (\ref{darkenergyeos}). In principle, a quite more general equation of state could be used, including for example powers of the Hubble rate, and thus including viscosity effects. We chose however the simplest of all the choices in order to simplify the calculations, however the outcomes of this study would be phenomenologically similar, namely the singularities would not appear in the loop quantum theory. In this section we shall investigate the phenomenological implications of the EoS (\ref{darkenergyeos}), in the context of LQC. Particularly, we shall be interested in observational data and constraints imposed on the luminosity distance moduli of Type IA supernovae, which we take from the Supernova Cosmology Project \cite{Amanullah:2010vv}, and also to Baryon Acoustic Oscillations, quantified by the phenomenological parameter $\mathcal{A}$ \cite{Eisenstein:2005su,Shafieloo:2012rs}. We shall adopt the notation and analysis of \cite{Astashenok:2012iy} which we extend in the context of LQC. Also we shall compare the classical theory with the LQC theory, and we discuss the types of singularities that occur in the single fluid description of the classical theory, for the EoS (\ref{darkenergyeos}).

Let us start off our analysis with the comparison of the classical theory with the LQC theory. In the case of the classical theory, an EoS of the form (\ref{darkenergyeos}) corresponding to an interacting dark energy-dark matter fluids system filling the Universe, unavoidably leads to finite-time singularities, and particularly to Type III singularities, as was shown in Ref. \cite{Nojiri:2005sx}. In the single dark energy fluid case of LQC, the singularities are erased \cite{Sami:2006wj}, and the same applies to our case, in which three fluids are present. Therefore, the LQC framework provides a singularity-free cosmological description of the dark energy era. We gather the results in Table \ref{table1}.
\begin{table*}[h]
\small \caption{\label{table1}Singularity Occurrence in Classical and LQC Description for the EoS $p_d=-\rho_d-A\kappa^4\rho_d^2$}
\begin{tabular}{@{}crrrrrrrrrrr@{}}
\tableline \tableline \tableline
 Classical Case: & Interacting Dark Energy-Dark Matter Fluids Physical & Type III Singularity.
\\\tableline
LQC case: & Single Dark Energy Fluid &  No finite-time singularity\\
\\\tableline
LQC case: & Three Cosmological Fluids &  No finite-time singularity\\
\\\tableline
 \tableline
\end{tabular}
\end{table*}
Now let us proceed to the phenomenological implications of the dark energy EoS (\ref{darkenergyeos}), in the context of LQC. Consider first the data on the luminosity distance modulus of Type IA supernovae, which if the supernova is at a redshift $z=a_0/a-1$, the distance modulus is equal to,
\begin{equation}\label{distancemodulus}
\mu (z)=\mathrm{const}+5\log D(z)\, ,
\end{equation}
where $D(z)$ is the luminosity distance. The latter is given by the following formula,
\begin{equation}\label{luminositydistance}
D(z)=\frac{c}{H_0}(1+z)\int_0^z h^{-1}(z)dz\, ,
\end{equation}
where the function $h(z)$ stands for,
\begin{equation}\label{hz}
h(z)^2=\frac{\rho_d(z)}{\rho_0}\, ,
\end{equation}
and $\rho_d(z)$ is the energy density of the dark energy fluid, and  $\rho_0$ is the total energy density at present time. Also, $c$ in Eq. (\ref{luminositydistance}) is the speed of light and $H_0$ is the Hubble rate at present time, which is approximately $H_0=72km/s$. Let us perform a statistical analysis for some sample redshifts, so we use the $\chi_{SN}^2$ statistics for the luminosity distance moduli, and we calculate $\chi_{SN}^2$, which is defined as,
\begin{equation}\label{chisquare}
\chi_{SN}^2=\frac{(\mu_{th}-\mu_{obs})^2}{\sigma_{\mu}^2}\, ,
\end{equation}
where $\mu_{th}$ is the theoretical prediction, $\mu_{obs}$ is the observed value for the luminosity distance modulus, and $\sigma_{\mu}$ is the standard deviation. Also the total $\chi_{SN}$ for a data set of redshifts, is equal to the sum of the corresponding $\chi_{SN}^2$ values for each redshift. In order to proceed it is vital to have the energy density of the dark energy fluid as a function of the redshift, and this can be found by using the continuity equation for the dark energy fluid, in LQC, which also contains a non-trivial interaction term $Q$. By assuming that the interaction term has the form $Q=3Hc_2\rho_d$ for simplicity, from the continuity equation we obtain,
\begin{equation}\label{contex1}
d\rho_d=(3\alpha-3c_2\rho_d)\frac{da}{a}\, ,
\end{equation}
where we have set $\alpha=A\kappa^4$. Upon integration we obtain,
\begin{equation}\label{pheneqn2}
\frac{a}{a_0}=\rho_{D0}\left ( \frac{c_2-\alpha\rho_d^2}{\rho_d}\right)^{\frac{1}{c_2}}\, ,
\end{equation}
where $\rho_{D0}$ is the current energy density of the dark energy fluid. By using the relation $z=a_0/a-1$, we can express the dark energy density as a function of the redshift, which is,
\begin{equation}\label{darkenergyeos}
 \rho_d(z)=\frac{\sqrt{4 \alpha  c_2 \rho_{D0}^2+(z+1)^{6 c_2}}-(z+1)^{3 c_2}}{2 \alpha  \rho_{D0}}\, ,
\end{equation}
and by using the equalities $\rho_{D0}=\Omega_{D0}\rho_0=3\Omega_{D0}H_0^2$, where $\Omega_{D0}\simeq 0.72$, we get,
\begin{equation}\label{darkenergyeos1}
 \rho_d(z)=\frac{\sqrt{36 \alpha  c_2 H_0^4 \Omega_{D0}^2+(z+1)^{6 c_2}}-(z+1)^{3 c_2}}{6 \alpha  H_0^2 \Omega_{D0}}\, .
\end{equation}
By using the above, we can perform a data analysis on the luminosity distance moduli, and obtain the $\chi_{SN}$ for a set of redshifts. The phenomenologically viable cases are plenty, as it proves, since the presence of two free variables, namely $c_2$ and $\alpha$, allows for many possible viable cases. We gather the results in Table \ref{table2}, and we note that the results should be compared with the best fit of the observational data, which correspond to the $\Lambda$CDM model, which yield $\chi_{SN}^2=347.06$ for $\Omega_{D0}\simeq 0.72$. As it can be seen in Table \ref{table2}, there are many possible values of $c_2$ and $\alpha$ which yield phenomenologically acceptable results.
\begin{table*}[h]
\small \caption{\label{table2}The values of $\chi_{SN}^2$ for various $(c_2,\alpha)$}
\begin{tabular}{@{}crrrrrrrrrrr@{}}
\tableline \tableline \tableline
 $(c_2,\alpha)$ & $\chi_{SN}^2$
\\\tableline
$(10^{-8},1000)$& 376.353
\\\tableline
$(2\times 10^{-8},1000)$ & 364.748
\\\tableline
$(6\times 10^{-8},1000)$ &346.798
\\
\tableline
$(7\times 10^{-8},1000)$  &344.321 \\
\tableline
 $(10^{-8},100)$ & 339.848\\
 \tableline
 $(10^{-8},120)$ &342.626 \\
 \tableline
 $(10^{-8},140)$ & 344.991\\
 \tableline
 $(10^{-8},170)$ & 347.992\\
 \tableline
\end{tabular}
\end{table*}
In order to have a more concrete qualitative idea on how the parameters $c_2$ and $\alpha$ affect the phenomenological implications of the dark energy model at hand, in Fig. (\ref{plot1}), we present the contour plot of $\chi_{SN}^2$ for various $(c_2\times 10^{-9},\alpha)$ values.
\begin{figure}[h]
\centering
\includegraphics[width=20pc]{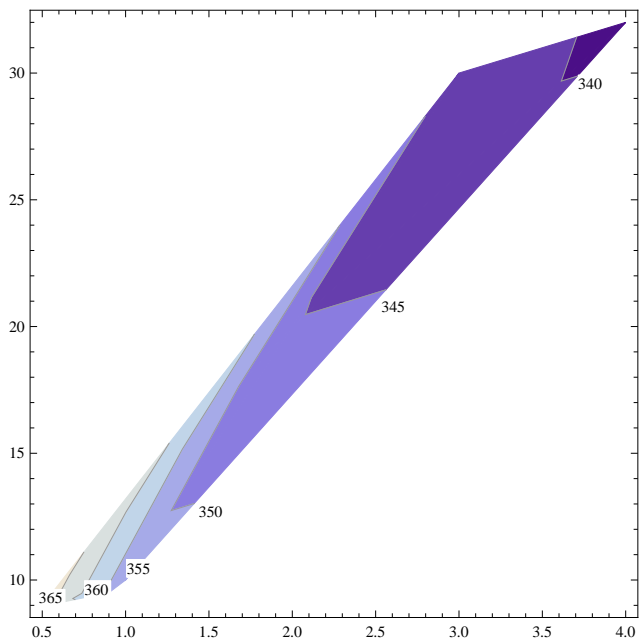}
\caption{{\it{Contour plot of $\chi_{SN}^2$ for various $(c_2\times 10^{-9},\alpha)$ values. The horizontal axis corresponds to $c_2$ and the vertical axis to $\alpha$.}}} \label{plot1}
\end{figure}
At it can be seen from Fig. (\ref{plot1}), the phenomenologically acceptable values of $\chi_{SN}^2$, are obtained for a wide range of the parameters $(c_2\times 10^{-9},\alpha)$, however it should be mentioned that this feature is strongly model dependent. Nevertheless, a viable phenomenology can be obtained by appropriately choosing the dark energy EoS.

Finally, let us briefly discuss the implications of the model at hand on the  Baryon Acoustic Oscillations (BAO). It was suggested \cite{Eisenstein:2005su,Shafieloo:2012rs} that the measurement related to BAO should be related t the large scale correlation function at $100h^{-1}$Mpc separations, using red galaxies. The most appropriate quantity to measure in this case, is the parameter $\mathcal{A}$, which is equal to,
\begin{equation}\label{alphabao}
\mathcal{A}=\sqrt{\Omega_{m0}}h(z_0)^{-1}\Big{[}\frac{1}{z_0}\int_0^{z_0}h^{-1}(z)d\,z\Big{]}^{2/3}\, ,
\end{equation}
where $z_0=0.35$ and $\Omega_{m0}=0.24$. The observational constraints on $\mathcal{A}$ are $\mathcal{A}=0.469\pm 0.017$, and by studying the parameter space of the model at hand, it is easily seen that the theoretical prediction for the BAO parameter is way too large in comparison to the observational constraint. For example, if $\alpha=1.8$ and $c_2=3.1$, we get $\mathcal{A}=53.3351$, and $\chi_{SN}^2=347.57097$. Thus the BAO constraint is not satisfied, although the observational constraints on the luminosity distance modulus is respected. In principle, it is possible to modify the dark energy EoS, in order to obtain validity with the observational data, for example by adding the term $1/\rho_d^{\alpha}$, a Chaplygin type gas generalization, the dark energy EoS would lead to BAO compatible with observations \cite{Biesiada:2004td}. However, a fractional EoS would make the dynamical system analysis more involved, so we did not study this case, for the sake of simplicity. In principle though, such a study is feasible.

As we already noted above, the model can be appropriately chosen so the resulting phenomenology is viable, in this paper we chose the simplest case of a generalized EoS for the shake of simplicity of the calculations, but we believe that if more complicated EoS's are used, the qualitative results should be the same, and in effect finite-time singularities would occur in three fluid models in the context of LQC.

Also a more appealing treatment should require to use the reduced $\chi^2_{SN}$ index, denoted as $r$, which is defined as the fraction of $\chi^2_{SN}$ over the degrees of freedom, with the latter being the total number of the supernovae used minus the free parameters of the theory. If $r\simeq 1$ then the model is acceptable observationally, and if $r\gg 1$, the model is not so appealing. Also if $r\ll 1$, the statistical errors could be large. In this study we confined ourselves on the comparison of the $\chi^2_{SN}$ obtained for the model at hand, with the one obtained from the $\Lambda$CDM model, however a more correct treatment should use the reduced $\chi^2$ index. However we did not go into this analysis due to the fact that this is a more focused study on the observational implications of such an EoS used, which was not our main aim. In principle the resulting picture would be comparable to the $\Lambda$CDM picture, since we use only two extra parameters.

\section{Conclusions}

In this paper we investigated the finite-time singularity structure of the cosmological dynamical system corresponding to  three cosmic fluids in the context of LQC. The three fluids consist of the two dark sector fluids, namely the dark energy and dark matter fluids, which were assumed to have a non-trivial interaction term, and in addition of a baryonic fluid which is not interacting with the rest of the two fluids. For the dark energy fluid we assumed a quite general equation of state, in order to take into account all the possible scenarios for this sector. The approach we adopted in order to investigate the existence of finite-time cosmological singularities, was fully analytic and relied upon a rigid theorem of dynamical systems, which enabled us to explicitly answer the question whether finite-time singularities occur. As we demonstrated, the LQC cosmological system has surely general non-singular solutions, and with general, it is meant that these solutions correspond to a very general set of initial conditions. In addition, we demonstrated that singular solutions also may be found, if some requirements are met, however these solutions correspond to a limited set of initial conditions, and therefore are not general solutions. Moreover, we should note that we discussed the difference of a physical finite-time singularity and of a singularity corresponding to a dynamical system. This discrimination is vital, and some overlap between the two systems exists, if the variables of the dynamical system are such, that they allow for physical conclusions to be made. In the case at hand, the form of the variables of the dynamical system allowed to demonstrate that the three fluids system in the context of LQC will never develop Type I (Big Rip) or Type III singularities. Thus the LQC context in this case too, actually removes the crushing type finite-time singularities from the cosmological theory, and the major contribution of this work is that we were able to prove this analytically, in terms of a rigid mathematical framework, without invoking numerical treatments. In order to achieve this, the dynamical system variables were chosen in such a way so that the resulting dynamical system is an autonomous polynomial dynamical system. Finally, in order to have a quantitative idea of what are the new physics that LQC introduces in the three fluids system, we studied the dark energy EoS and we performed a comparison of the three fluids case with the single dark energy fluid case. In the classical single dark energy fluid case, finite-time singularities could not be avoided, and also in the LQC case of a single dark energy fluid, no singularities occur \cite{Sami:2006wj}, at least for the dark energy EoS we used. As we analytically showed with our work, the same applies for the three fluids LQC case and we provided sufficient proofs in an analytic way.

In principle, more general dark energy equations of state can be used, however we focused on a simple generalized form of the $p_d=-\rho_d$ case, which was $p_d=-\rho_d-A\rho_d^2$. In this case the calculations were proven simpler, and the dynamical system could be handled analytically by using the mathematical theorem proved in \cite{goriely}. We should note that more general integer powers for the energy density the equation of state can be used, namely $p_d=-\rho_d-A\rho_d^n$, $n>2$, or even including powers of the Hubble rate, in order to include some higher order viscosity contributions, however no fractional powers of the energy density are allowed, that is when $n<1$. This case would require another approach, since the resulting dynamical system would not be polynomial type, and hence no concrete answer on the finite-time singularity structure of the cosmological dynamical system can be made.

Finally, we need to discuss a question that naturally springs to mind, namely whether the non-occurrence of finite-time singularities could have been anticipated by the inherent structure of LQC. This is a challenging theoretical question, it is possible that the fact of having a modified and constrained Friedmann equation, may actually constrain the dynamics of the Universe so that finite-time singularities do not occur. This is not an easy issue to discuss theoretically, however all the studies in LQC finite-time singularities, strongly indicate that these disappear from the theory, and in addition the present study indicates that this occurs even in the case of the presence of various cosmological fluids by using the dynamical systems approach. Therefore this is a indirect proof that the inherent structure of LQC crucially affects the development of finite-time singularities.

\section*{Acknowledgments}

This work is supported by MINECO (Spain), FIS2016-76363-P (S.D.O) and by PHAROS-COST action No: CA16214 (S.D. Odintsov and V.K. Oikonomou). Also V. Oikonomou is grateful to M. Plionis for helpful discussions on the $\chi^2$ statistics.


\begin{thebibliography}{99}




\bibitem{reviews1}
 S.~Nojiri, S.~D.~Odintsov and V.~K.~Oikonomou,
  Phys.\ Rept.\  {\bf 692} (2017) 1
  doi:10.1016/j.physrep.2017.06.001
  [arXiv:1705.11098 [gr-qc]].

\bibitem{reviews2}

S. Nojiri, S.D. Odintsov,
   Phys.\ Rept.\  {\bf 505}, 59 (2011);


  \bibitem{reviews3}
S. Nojiri, S.D. Odintsov,
  eConf {\bf C0602061}, 06 (2006)
  [Int.\ J.\ Geom.\ Meth.\ Mod.\ Phys.\  {\bf 4}, 115 (2007)].


   \bibitem{reviews4}
 S. Capozziello, M. De Laurentis,
   Phys.\ Rept.\  {\bf 509}, 167 (2011);\\
 V.~Faraoni and S.~Capozziello,
  Fundam.\ Theor.\ Phys.\  {\bf 170} (2010).
  doi:10.1007/978-94-007-0165-6



\bibitem{reviews5}

A.~de la Cruz-Dombriz and D.~Saez-Gomez,
  Entropy {\bf 14} (2012) 1717
  doi:10.3390/e14091717
  [arXiv:1207.2663 [gr-qc]].

\bibitem{reviews6}

G.~J.~Olmo,
  Int.\ J.\ Mod.\ Phys.\ D {\bf 20} (2011) 413
  doi:10.1142/S0218271811018925
  [arXiv:1101.3864 [gr-qc]].



\bibitem{LQC1}


A.~Ashtekar and P.~Singh,
Class.\ Quant.\ Grav.\  {\bf 28} (2011) 213001
[arXiv:1108.0893 [gr-qc]]



\bibitem{LQC3} A.~Ashtekar, T.~Pawlowski and P.~Singh,
  Phys.\ Rev.\ Lett.\  {\bf 96} (2006) 141301
  [gr-qc/0602086].


\bibitem{LQC4} A.~Ashtekar, T.~Pawlowski and P.~Singh,
  Phys.\ Rev.\ D {\bf 73} (2006) 124038
  [gr-qc/0604013].


\bibitem{LQC5}  A.~Ashtekar, T.~Pawlowski and P.~Singh,
  Phys.\ Rev.\ D {\bf 74} (2006) 084003
  [gr-qc/0607039].







\bibitem{Salo:2016dsr}
  L.~Areste Salo, J.~Amoros and J.~de Haro,
  Class.\ Quant.\ Grav.\  {\bf 34} (2017) no.23,  235001
  doi:10.1088/1361-6382/aa9311
  [arXiv:1612.05480 [gr-qc]].



\bibitem{Xiong:2007cn}
  H.~H.~Xiong, T.~Qiu, Y.~F.~Cai and X.~Zhang,
  Mod.\ Phys.\ Lett.\ A {\bf 24} (2009) 1237
  doi:10.1142/S0217732309030667
  [arXiv:0711.4469 [hep-th]].

\bibitem{Amoros:2014tha}
  J.~Amoros, J.~de Haro and S.~D.~Odintsov,
  Phys.\ Rev.\ D {\bf 89} (2014) no.10,  104010
  doi:10.1103/PhysRevD.89.104010
  [arXiv:1402.3071 [gr-qc]].


\bibitem{Cai:2014zga}
  Y.~F.~Cai and E.~Wilson-Ewing,
  JCAP {\bf 1403} (2014) 026
  doi:10.1088/1475-7516/2014/03/026
  [arXiv:1402.3009 [gr-qc]].


\bibitem{deHaro:2014kxa}
  J.~de Haro and J.~Amoros,
  JCAP {\bf 1408} (2014) 025
  doi:10.1088/1475-7516/2014/08/025
  [arXiv:1403.6396 [gr-qc]].





\bibitem{hawkingpenrose}
S.~W.~Hawking and R.~Penrose,
Proc.\ Roy.\ Soc.\ Lond.\ A {\bf 314} (1970) 529.



\bibitem{Santos:2007bs}
  J.~Santos, J.~S.~Alcaniz, M.~J.~Reboucas and F.~C.~Carvalho,
  Phys.\ Rev.\ D {\bf 76} (2007) 083513
  doi:10.1103/PhysRevD.76.083513
  [arXiv:0708.0411 [astro-ph]].





\bibitem{Nojiri:2005sx}
  S.~Nojiri, S.~D.~Odintsov and S.~Tsujikawa,
  Phys.\ Rev.\ D {\bf 71} (2005) 063004
  doi:10.1103/PhysRevD.71.063004
  [hep-th/0501025].




\bibitem{Sami:2006wj}
  M.~Sami, P.~Singh and S.~Tsujikawa,
  Phys.\ Rev.\ D {\bf 74} (2006) 043514
  doi:10.1103/PhysRevD.74.043514
  [gr-qc/0605113].


\bibitem{goriely}

 A. Goriely, C Hyde, Journal of Differential Equations 161, 422-448 (2000)\\ https://doi.org/10.1006/jdeq.1999.3688





\bibitem{Oikonomou:2006mh}
V.~K.~Oikonomou, J.~D.~Vergados and C.~C.~Moustakidis,
  Nucl.\ Phys.\ B {\bf 773} (2007) 19
  doi:10.1016/j.nuclphysb.2007.03.014
  [hep-ph/0612293].






\bibitem{Barrow:1994nx}
  J.~D.~Barrow and J.~P.~Mimoso,
  Phys.\ Rev.\ D {\bf 50} (1994) 3746.
  doi:10.1103/PhysRevD.50.3746


\bibitem{Tsagas:1998jm}
  C.~G.~Tsagas and J.~D.~Barrow,
  Class.\ Quant.\ Grav.\  {\bf 15} (1998) 3523
  doi:10.1088/0264-9381/15/11/016
  [gr-qc/9803032].


\bibitem{HipolitoRicaldi:2009je}
  W.~S.~Hipolito-Ricaldi, H.~E.~S.~Velten and W.~Zimdahl,
  JCAP {\bf 0906} (2009) 016
  doi:10.1088/1475-7516/2009/06/016
  [arXiv:0902.4710 [astro-ph.CO]].


\bibitem{Gorini:2005nw}
  V.~Gorini, A.~Kamenshchik, U.~Moschella, V.~Pasquier and A.~Starobinsky,
  Phys.\ Rev.\ D {\bf 72} (2005) 103518
  doi:10.1103/PhysRevD.72.103518
  [astro-ph/0504576].



\bibitem{Kremer:2003vs}
  G.~M.~Kremer,
  Phys.\ Rev.\ D {\bf 68} (2003) 123507
  doi:10.1103/PhysRevD.68.123507
  [gr-qc/0309111].


\bibitem{Carturan:2002si}
  D.~Carturan and F.~Finelli,
  Phys.\ Rev.\ D {\bf 68} (2003) 103501
  doi:10.1103/PhysRevD.68.103501
  [astro-ph/0211626].


\bibitem{Buchert:2001sa}
  T.~Buchert,
  Gen.\ Rel.\ Grav.\  {\bf 33} (2001) 1381
  doi:10.1023/A:1012061725841
  [gr-qc/0102049].


\bibitem{Hwang:2001fb}
  J.~c.~Hwang and H.~Noh,
  Class.\ Quant.\ Grav.\  {\bf 19} (2002) 527
  doi:10.1088/0264-9381/19/3/308
  [astro-ph/0103244].

\bibitem{Cruz:2011zza}
  N.~Cruz, S.~Lepe and F.~Pena,
  Phys.\ Lett.\ B {\bf 699} (2011) 135.
  doi:10.1016/j.physletb.2011.03.049



\bibitem{Oikonomou:2017mlk}
  V.~K.~Oikonomou,
  Int.\ J.\ Mod.\ Phys.\ D {\bf 26} (2017) no.10,  1750110
  doi:10.1142/S0218271817501103
  [arXiv:1703.09009 [gr-qc]].


\bibitem{Brevik:2017msy}
  I.~Brevik, O.~Gron, J.~de Haro, S.~D.~Odintsov and E.~N.~Saridakis,
  Int.\ J.\ Mod.\ Phys.\ D {\bf 26} (2017) no.14,  1730024
  doi:10.1142/S0218271817300245
  [arXiv:1706.02543 [gr-qc]].

\bibitem{Nojiri:2005sr}
  S.~Nojiri and S.~D.~Odintsov,
  Phys.\ Rev.\ D {\bf 72} (2005) 023003
  doi:10.1103/PhysRevD.72.023003
  [hep-th/0505215].



\bibitem{Capozziello:2006dj}
  S.~Capozziello, S.~Nojiri, S.~D.~Odintsov and A.~Troisi,
  Phys.\ Lett.\ B {\bf 639} (2006) 135
  doi:10.1016/j.physletb.2006.06.034
  [astro-ph/0604431].


\bibitem{Nojiri:2006zh}
  S.~Nojiri and S.~D.~Odintsov,
  Phys.\ Lett.\ B {\bf 639} (2006) 144
  doi:10.1016/j.physletb.2006.06.065
  [hep-th/0606025].

\bibitem{Elizalde:2009gx}
  E.~Elizalde and D.~Saez-Gomez,
  Phys.\ Rev.\ D {\bf 80} (2009) 044030
  doi:10.1103/PhysRevD.80.044030
  [arXiv:0903.2732 [hep-th]].









\bibitem{Elizalde:2017dmu}
  E.~Elizalde and M.~Khurshudyan,
  arXiv:1711.01143 [gr-qc].




\bibitem{Brevik:2016kuy}
  I.~Brevik and A.~V.~Timoshkin,
  Int.\ J.\ Geom.\ Meth.\ Mod.\ Phys.\  {\bf 14} (2017) no.04,  1750061
  doi:10.1142/S021988781750061X
  [arXiv:1612.06689 [gr-qc]].


\bibitem{Balakin:2012ee}
  A.~B.~Balakin and V.~V.~Bochkarev,
  Phys.\ Rev.\ D {\bf 87} (2013) no.2,  024006
  doi:10.1103/PhysRevD.87.024006
  [arXiv:1212.4094 [gr-qc]].


\bibitem{Zimdahl:1998rx}
  W.~Zimdahl and A.~B.~Balakin,
  Class.\ Quant.\ Grav.\  {\bf 15} (1998) 3259
  doi:10.1088/0264-9381/15/10/026
  [gr-qc/9807078].




\bibitem{Kunz:2007rk}
  M.~Kunz,
  Phys.\ Rev.\ D {\bf 80} (2009) 123001
  doi:10.1103/PhysRevD.80.123001
  [astro-ph/0702615].





\bibitem{Gondolo:2002fh}
  P.~Gondolo and K.~Freese,
  Phys.\ Rev.\ D {\bf 68} (2003) 063509
  doi:10.1103/PhysRevD.68.063509
  [hep-ph/0209322].


\bibitem{Farrar:2003uw}
  G.~R.~Farrar and P.~J.~E.~Peebles,
  Astrophys.\ J.\  {\bf 604} (2004) 1
  doi:10.1086/381728
  [astro-ph/0307316].



\bibitem{Cai:2004dk}
  R.~G.~Cai and A.~Wang,
  JCAP {\bf 0503} (2005) 002
  doi:10.1088/1475-7516/2005/03/002
  [hep-th/0411025].


\bibitem{Bamba:2012cp}
  K.~Bamba, S.~Capozziello, S.~Nojiri and S.~D.~Odintsov,
  Astrophys.\ Space Sci.\  {\bf 342} (2012) 155
  doi:10.1007/s10509-012-1181-8
  [arXiv:1205.3421 [gr-qc]].



\bibitem{Guo:2004xx}
  Z.~K.~Guo, R.~G.~Cai and Y.~Z.~Zhang,
  JCAP {\bf 0505} (2005) 002
  doi:10.1088/1475-7516/2005/05/002
  [astro-ph/0412624].



\bibitem{Wang:2006qw}
  B.~Wang, J.~Zang, C.~Y.~Lin, E.~Abdalla and S.~Micheletti,
  Nucl.\ Phys.\ B {\bf 778} (2007) 69
  doi:10.1016/j.nuclphysb.2007.04.037
  [astro-ph/0607126].



\bibitem{Bertolami:2007zm}
  O.~Bertolami, F.~Gil Pedro and M.~Le Delliou,
  Phys.\ Lett.\ B {\bf 654} (2007) 165
  doi:10.1016/j.physletb.2007.08.046
  [astro-ph/0703462 [ASTRO-PH]].





\bibitem{He:2008tn}
  J.~H.~He and B.~Wang,
  JCAP {\bf 0806} (2008) 010
  doi:10.1088/1475-7516/2008/06/010
  [arXiv:0801.4233 [astro-ph]].



\bibitem{Valiviita:2008iv}
  J.~Valiviita, E.~Majerotto and R.~Maartens,
  JCAP {\bf 0807} (2008) 020
  doi:10.1088/1475-7516/2008/07/020
  [arXiv:0804.0232 [astro-ph]].


\bibitem{Jackson:2009mz}
  B.~M.~Jackson, A.~Taylor and A.~Berera,
  Phys.\ Rev.\ D {\bf 79} (2009) 043526
  doi:10.1103/PhysRevD.79.043526
  [arXiv:0901.3272 [astro-ph.CO]].


\bibitem{Jamil:2009eb}
  M.~Jamil, E.~N.~Saridakis and M.~R.~Setare,
  Phys.\ Rev.\ D {\bf 81} (2010) 023007
  doi:10.1103/PhysRevD.81.023007
  [arXiv:0910.0822 [hep-th]].




\bibitem{He:2010im}
  J.~H.~He, B.~Wang and E.~Abdalla,
  Phys.\ Rev.\ D {\bf 83} (2011) 063515
  doi:10.1103/PhysRevD.83.063515
  [arXiv:1012.3904 [astro-ph.CO]].


\bibitem{Bolotin:2013jpa}
  Y.~L.~Bolotin, A.~Kostenko, O.~A.~Lemets and D.~A.~Yerokhin,
  Int.\ J.\ Mod.\ Phys.\ D {\bf 24} (2014) no.03,  1530007
  doi:10.1142/S0218271815300074
  [arXiv:1310.0085 [astro-ph.CO]].


\bibitem{Costa:2013sva}
  A.~A.~Costa, X.~D.~Xu, B.~Wang, E.~G.~M.~Ferreira and E.~Abdalla,
  Phys.\ Rev.\ D {\bf 89} (2014) no.10,  103531
  doi:10.1103/PhysRevD.89.103531
  [arXiv:1311.7380 [astro-ph.CO]].




\bibitem{Boehmer:2008av}
  C.~G.~Boehmer, G.~Caldera-Cabral, R.~Lazkoz and R.~Maartens,
  Phys.\ Rev.\ D {\bf 78} (2008) 023505
  doi:10.1103/PhysRevD.78.023505
  [arXiv:0801.1565 [gr-qc]].



\bibitem{Li:2010ju}
  S.~Li and Y.~Ma,
  Eur.\ Phys.\ J.\ C {\bf 68} (2010) 227
  doi:10.1140/epjc/s10052-010-1338-y
  [arXiv:1004.4350 [astro-ph.CO]].


\bibitem{Yang:2017zjs}
  W.~Yang, S.~Pan and J.~D.~Barrow,
  arXiv:1706.04953 [astro-ph.CO].












\bibitem{Boehmer:2014vea}
  C.~G.~Boehmer and N.~Chan,
  doi:10.1142/9781786341044.0004
  arXiv:1409.5585 [gr-qc].




\bibitem{Bohmer:2010re}
  C.~G.~Boehmer, T.~Harko and S.~V.~Sabau,
  Adv.\ Theor.\ Math.\ Phys.\  {\bf 16} (2012) no.4,  1145
  doi:10.4310/ATMP.2012.v16.n4.a2
  [arXiv:1010.5464 [math-ph]].





\bibitem{Goheer:2007wu}
  N.~Goheer, J.~A.~Leach and P.~K.~S.~Dunsby,
  Class.\ Quant.\ Grav.\  {\bf 24} (2007) 5689
  doi:10.1088/0264-9381/24/22/026
  [arXiv:0710.0814 [gr-qc]].





\bibitem{Leon:2014yua}
  G.~Leon and E.~N.~Saridakis,
  JCAP {\bf 1504} (2015) no.04,  031
  doi:10.1088/1475-7516/2015/04/031
  [arXiv:1501.00488 [gr-qc]].





\bibitem{Leon:2010pu}
  G.~Leon and E.~N.~Saridakis,
  Class.\ Quant.\ Grav.\  {\bf 28} (2011) 065008
  doi:10.1088/0264-9381/28/6/065008
  [arXiv:1007.3956 [gr-qc]].


\bibitem{deSouza:2007zpn}
  J.~C.~C.~de Souza and V.~Faraoni,
  Class.\ Quant.\ Grav.\  {\bf 24} (2007) 3637
  doi:10.1088/0264-9381/24/14/006
  [arXiv:0706.1223 [gr-qc]].

\bibitem{Giacomini:2017yuk}
  A.~Giacomini, S.~Jamal, G.~Leon, A.~Paliathanasis and J.~Saavedra,
  Phys.\ Rev.\ D {\bf 95} (2017) no.12,  124060
  doi:10.1103/PhysRevD.95.124060
  [arXiv:1703.05860 [gr-qc]].


\bibitem{Kofinas:2014aka}
  G.~Kofinas, G.~Leon and E.~N.~Saridakis,
  Class.\ Quant.\ Grav.\  {\bf 31} (2014) 175011
  doi:10.1088/0264-9381/31/17/175011
  [arXiv:1404.7100 [gr-qc]].


\bibitem{Leon:2012mt}
  G.~Leon and E.~N.~Saridakis,
  JCAP {\bf 1303} (2013) 025
  doi:10.1088/1475-7516/2013/03/025
  [arXiv:1211.3088 [astro-ph.CO]].



\bibitem{Gonzalez:2006cj}
  T.~Gonzalez, G.~Leon and I.~Quiros,
  Class.\ Quant.\ Grav.\  {\bf 23} (2006) 3165
  doi:10.1088/0264-9381/23/9/025
  [astro-ph/0702227].



\bibitem{Alho:2016gzi}
  A.~Alho, S.~Carloni and C.~Uggla,
  JCAP {\bf 1608} (2016) no.08,  064
  doi:10.1088/1475-7516/2016/08/064
  [arXiv:1607.05715 [gr-qc]].


\bibitem{Biswas:2015cva}
  S.~K.~Biswas and S.~Chakraborty,
  Int.\ J.\ Mod.\ Phys.\ D {\bf 24} (2015) no.07,  1550046
  doi:10.1142/S0218271815500467
  [arXiv:1504.02431 [gr-qc]].


\bibitem{Muller:2014qja}
  D.~Müller, V.~C.~de Andrade, C.~Maia, M.~J.~Rebouças and A.~F.~F.~Teixeira,
  Eur.\ Phys.\ J.\ C {\bf 75} (2015) no.1,  13
  doi:10.1140/epjc/s10052-014-3227-2
  [arXiv:1405.0768 [astro-ph.CO]].





\bibitem{Mirza:2014nfa}
  B.~Mirza and F.~Oboudiat,
  Int.\ J.\ Geom.\ Meth.\ Mod.\ Phys.\  {\bf 13} (2016) no.09,  1650108
  doi:10.1142/S0219887816501085
  [arXiv:1412.6640 [gr-qc]].


\bibitem{Rippl:1995bg}
  S.~Rippl, H.~van Elst, R.~K.~Tavakol and D.~Taylor,
  Gen.\ Rel.\ Grav.\  {\bf 28} (1996) 193
  doi:10.1007/BF02105423
  [gr-qc/9511010].


\bibitem{Ivanov:2011vy}
  M.~M.~Ivanov and A.~V.~Toporensky,
  Grav.\ Cosmol.\  {\bf 18} (2012) 43
  doi:10.1134/S0202289312010100
  [arXiv:1106.5179 [gr-qc]].


\bibitem{Khurshudyan:2016qox}
  M.~Khurshudyan,
  Int.\ J.\ Geom.\ Meth.\ Mod.\ Phys.\  {\bf 14} (2016) no.03,  1750041.
  doi:10.1142/S0219887817500414


\bibitem{Boko:2016mwr}
  R.~D.~Boko, M.~J.~S.~Houndjo and J.~Tossa,
  Int.\ J.\ Mod.\ Phys.\ D {\bf 25} (2016) no.10,  1650098
  doi:10.1142/S021827181650098X
  [arXiv:1605.03404 [gr-qc]].


\bibitem{Odintsov:2017icc}
  S.~D.~Odintsov, V.~K.~Oikonomou and P.~V.~Tretyakov,
  Phys.\ Rev.\ D {\bf 96} (2017) no.4,  044022
  doi:10.1103/PhysRevD.96.044022
  [arXiv:1707.08661 [gr-qc]].


\bibitem{Odintsov:2017tbc}
  S.~D.~Odintsov and V.~K.~Oikonomou,
  Phys.\ Rev.\ D {\bf 96} (2017) no.10,  104049
  doi:10.1103/PhysRevD.96.104049
  [arXiv:1711.02230 [gr-qc]].


\bibitem{Oikonomou:2017ppp}
  V.~K.~Oikonomou,
  arXiv:1711.03389 [gr-qc].




\bibitem{Odintsov:2015wwp}
  S.~D.~Odintsov and V.~K.~Oikonomou,
  Phys.\ Rev.\ D {\bf 93} (2016) no.2,  023517
  doi:10.1103/PhysRevD.93.023517
  [arXiv:1511.04559 [gr-qc]].


\bibitem{Bahamonde:2017ize}
  S.~Bahamonde, C.~G.~Boehmer, S.~Carloni, E.~J.~Copeland, W.~Fang and N.~Tamanini,
  arXiv:1712.03107 [gr-qc].


\bibitem{Ganiou:2018dta}
  M.~G.~Ganiou, P.~H.~Logbo, M.~J.~S.~Houndjo and J.~Tossa,
  arXiv:1805.00332 [gr-qc].









\bibitem{Xiao:2010cy}
  K.~Xiao and J.~Y.~Zhu,
  Int.\ J.\ Mod.\ Phys.\ A {\bf 25} (2010) 4993
  doi:10.1142/S0217751X10050585
  [arXiv:1006.5377 [gr-qc]].


\bibitem{Zonunmawia:2017ofc}
  H.~Zonunmawia, W.~Khyllep, N.~Roy, J.~Dutta and N.~Tamanini,
  Phys.\ Rev.\ D {\bf 96} (2017) no.8,  083527
  doi:10.1103/PhysRevD.96.083527
  [arXiv:1708.07716 [gr-qc]].



\bibitem{Chen:2008ca}
  S.~Chen, B.~Wang and J.~Jing,
  Phys.\ Rev.\ D {\bf 78} (2008) 123503
  doi:10.1103/PhysRevD.78.123503
  [arXiv:0808.3482 [gr-qc]].



 \bibitem{barrowcotsakis}

  S.~Cotsakis and J.~D.~Barrow,
  J.\ Phys.\ Conf.\ Ser.\  {\bf 68} (2007) 012004
  doi:10.1088/1742-6596/68/1/012004
  [gr-qc/0608137].






\bibitem{abl03}
 A.Ashtekar, M. Bojowald and J.Lewandowski, Adv.Theor.Math.Phys.
{\bf 7}, 233-
268 (2003) [arXiv:0304074].


\bibitem{bojowald05}
  M. Bojowald, Living Rev. Rel.
{\bf 8}, 11 (2005) [arXiv:0601085].



\bibitem{Singh07}
 P. Singh, J.Phys.Conf.Ser. {\bf 140}, 012005 (2008) [arXiv:0901.1301].




\bibitem{he}  J. Haro, E. Elizalde, EPL
{\bf 89},
69001 (2010).


\bibitem{dmp}  P. Dzierzak, P. Malkiewicz and W. Piechocki, Phys. Rev.
{\bf D 80}, 104001 (2009)
[arXiv:0907.3436].


\bibitem{Martineau:2017sti}
  K.~Martineau, A.~Barrau and S.~Schander,
  Phys.\ Rev.\ D {\bf 95} (2017) no.8,  083507
  doi:10.1103/PhysRevD.95.083507
  [arXiv:1701.02703 [gr-qc]].

\bibitem{Barrau:2014maa}
  A.~Barrau, M.~Bojowald, G.~Calcagni, J.~Grain and M.~Kagan,
  JCAP {\bf 1505} (2015) no.05,  051
  doi:10.1088/1475-7516/2015/05/051
  [arXiv:1404.1018 [gr-qc]].
  
  
\bibitem{Agullo:2012sh}
  I.~Agullo, A.~Ashtekar and W.~Nelson,
  Phys.\ Rev.\ Lett.\  {\bf 109} (2012) 251301
  doi:10.1103/PhysRevLett.109.251301
  [arXiv:1209.1609 [gr-qc]].
  
  
\bibitem{Alesci:2016gub}
  E.~Alesci and F.~Cianfrani,
  Int.\ J.\ Mod.\ Phys.\ D {\bf 25} (2016) no.08,  1642005
  doi:10.1142/S0218271816420050
  [arXiv:1602.05475 [gr-qc]].
  
  
\bibitem{Martineau:2017tdx}
  K.~Martineau, A.~Barrau and J.~Grain,
  Int.\ J.\ Mod.\ Phys.\ D {\bf 27} (2018) no.07,  1850067
  doi:10.1142/S0218271818500670
  [arXiv:1709.03301 [gr-qc]].


\bibitem{Diener:2017lde}
  P.~Diener, A.~Joe, M.~Megevand and P.~Singh,
  Class.\ Quant.\ Grav.\  {\bf 34} (2017) no.9,  094004
  doi:10.1088/1361-6382/aa68b5
  [arXiv:1701.05824 [gr-qc]].






\bibitem{CalderaCabral:2008bx}
  G.~Caldera-Cabral, R.~Maartens and L.~A.~Urena-Lopez,
  Phys.\ Rev.\ D {\bf 79} (2009) 063518
  doi:10.1103/PhysRevD.79.063518
  [arXiv:0812.1827 [gr-qc]].


\bibitem{Pavon:2005yx}
  D.~Pavon and W.~Zimdahl,
  Phys.\ Lett.\ B {\bf 628} (2005) 206
  doi:10.1016/j.physletb.2005.08.134
  [gr-qc/0505020].

\bibitem{Quartin:2008px}
  M.~Quartin, M.~O.~Calvao, S.~E.~Joras, R.~R.~R.~Reis and I.~Waga,
  JCAP {\bf 0805} (2008) 007
  doi:10.1088/1475-7516/2008/05/007
  [arXiv:0802.0546 [astro-ph]].


\bibitem{Sadjadi:2006qp}
  H.~M.~Sadjadi and M.~Alimohammadi,
  Phys.\ Rev.\ D {\bf 74} (2006) 103007
  doi:10.1103/PhysRevD.74.103007
  [gr-qc/0610080].


\bibitem{Zimdahl:2005bk}
  W.~Zimdahl,
  Int.\ J.\ Mod.\ Phys.\ D {\bf 14} (2005) 2319
  doi:10.1142/S0218271805007784
  [gr-qc/0505056].








 \bibitem{barrowsudden}
J.~D.~Barrow,
Class.\ Quant.\ Grav.\  {\bf 21} (2004) L79 [gr-qc/0403084]. ;\\
S.~Nojiri and S.~D.~Odintsov,
  Phys.\ Lett.\ B {\bf 595} (2004) 1
  doi:10.1016/j.physletb.2004.06.060
  [hep-th/0405078].



\bibitem{barrowsudden1}
J.~D.~Barrow and C.~G.~Tsagas,
Class.\ Quant.\ Grav.\ {\bf 22}, 1563 (2005)
[arXiv:gr-qc/0411045]; \\
L.~Fernandez-Jambrina and
R.~Lazkoz,
Phys.\ Rev.\ D {\bf 70}, 121503 (2004) [arXiv:gr-qc/0410124];\\
M.~Bouhmadi-Lopez, P.~F.~Gonzalez-Diaz and P.~Martin-Moruno,
Phys.\ Lett.\ B {\bf 659}, 1 (2008) [arXiv:gr-qc/0612135];\\
J.~D.~Barrow and S.~Z.~W.~Lip,
arXiv:0901.1626 [gr-qc]; \\
M.~Bouhmadi-Lopez, Y.~Tavakoli and P.~V.~Moniz,
arXiv:0911.1428 [gr-qc].; \\
J.~D.~Barrow, A.~B.~Batista, J.~C.~Fabris, M.~J.~S.~Houndjo and
G.~Dito,
Phys.\ Rev.\ D {\bf 84} (2011) 123518
[arXiv:1110.1321 [gr-qc]];\\
M.~Bouhmadi-Lopez, C.~Kiefer and M.~Kramer,
  Phys.\ Rev.\ D {\bf 89} (2014) 6,  064016
  [arXiv:1312.5976 [gr-qc]]; \\
  M.~Bouhmadi-Lopez, P.~Chen and Y.~W.~Liu,
  Eur.\ Phys.\ J.\ C {\bf 73} (2013) 2546
  [arXiv:1302.6249 [gr-qc]].; \\
   A.~Balcerzak and M.~P.~Dabrowski,
  Phys.\ Rev.\ D {\bf 73} (2006) 101301
  [hep-th/0604034].





\bibitem{Odintsov:2015gba}
  S.~D.~Odintsov and V.~K.~Oikonomou,
  Phys.\ Rev.\ D {\bf 92} (2015) no.12,  124024
  doi:10.1103/PhysRevD.92.124024
  [arXiv:1510.04333 [gr-qc]].


\bibitem{Odintsov:2015jca}
  S.~D.~Odintsov and V.~K.~Oikonomou,
  Phys.\ Rev.\ D {\bf 92} (2015) no.2,  024058
  doi:10.1103/PhysRevD.92.024058
  [arXiv:1507.05273 [gr-qc]].



\bibitem{Barrow:2015ora}
J.~D.~Barrow and A.~A.~H.~Graham,
  Phys.\ Rev.\ D {\bf 91}, no. 8, 083513 (2015)
  [arXiv:1501.04090 [gr-qc]].




\bibitem{Nojiri:2015fra}
  S.~Nojiri, S.~D.~Odintsov and V.~K.~Oikonomou,
  Phys.\ Rev.\ D {\bf 91} (2015) no.8,  084059
  doi:10.1103/PhysRevD.91.084059
  [arXiv:1502.07005 [gr-qc]].




\bibitem{Oikonomou:2015qfh}
  V.~K.~Oikonomou,
  Int.\ J.\ Geom.\ Meth.\ Mod.\ Phys.\  {\bf 13} (2016) no.03,  1650033
  doi:10.1142/S021988781650033X
  [arXiv:1512.04095 [gr-qc]].



\bibitem{jost} Dynamical Systems, Examples of Complex Behaviour, Juergen Jost, Springer Universitext, 2005

\bibitem{wiggins} Stephen Wiggins, Introduction to Applied Nonlinear Dynamical Systems and
Chaos, Springer, New York, 2003

\bibitem{voyatzis} Introduction to non-Linear Dynamical Systems, George Vougiatzis, Eythimia Meletlidou, Kallipos (2015) (In Greek)


\bibitem{Amanullah:2010vv}
  R.~Amanullah {\it et al.},
  Astrophys.\ J.\  {\bf 716} (2010) 712
  doi:10.1088/0004-637X/716/1/712
  [arXiv:1004.1711 [astro-ph.CO]].




\bibitem{Eisenstein:2005su}
  D.~J.~Eisenstein {\it et al.} [SDSS Collaboration],
  Astrophys.\ J.\  {\bf 633} (2005) 560
  doi:10.1086/466512
  [astro-ph/0501171].


\bibitem{Shafieloo:2012rs}
  A.~Shafieloo, V.~Sahni and A.~A.~Starobinsky,
  Phys.\ Rev.\ D {\bf 86} (2012) 103527
  doi:10.1103/PhysRevD.86.103527
  [arXiv:1205.2870 [astro-ph.CO]].


\bibitem{Astashenok:2012iy}
  A.~V.~Astashenok and S.~D.~Odintsov,
  Phys.\ Lett.\ B {\bf 718} (2013) 1194
  doi:10.1016/j.physletb.2012.12.058
  [arXiv:1211.1888 [gr-qc]].


\bibitem{Biesiada:2004td}
  M.~Biesiada, W.~Godlowski and M.~Szydlowski,
  Astrophys.\ J.\  {\bf 622} (2005) 28
  doi:10.1086/427863
  [astro-ph/0403305].



\end{thebibliography}
\end{document}